\documentclass[%
 reprint,
 amsmath,amssymb,
 aps,
]{revtex4-2}

\usepackage{graphicx}
\usepackage{dcolumn}
\usepackage{bm}
\usepackage{hyperref}
\usepackage{graphicx}
\usepackage{braket}
\usepackage{amsfonts}
\usepackage{xcolor}
\newcommand{\abs}[1]{\left\vert#1\right\vert}

\begin{document}

\title{Theory of High-Temperature Superfluorescence in Hybrid Perovskite Thin Films.}
\author{B. D. Fainberg}
\email{fainberg@hit.ac.il }
\affiliation{H.I.T.-Holon Institute of Technology, 52 Golomb Street, POB 305 Holon 5810201, Israel}
\author{V. Al. Osipov}
\email{vladimir.al.osipov@gmail.com}
\affiliation{Institute of Advanced Study in Mathematics of Harbin Institute of Technology, 92 West Da Zhi Street, Harbin 150001, China}
\affiliation{Suzhou Research Institute of Harbin Institute of Technology, 500 South Guandu Road, Suzhou, 215104, China}
\date{\today }

\begin{abstract}
The recent discovery of high-temperature superfluorescence in hybrid perovskite thin films has opened new possibilities for harnessing macroscopic quantum phenomena in nanotechnology. This study aimed to elucidate the mechanism that enables high-temperature superfluorescence in these systems. The proposed model describes a quasi-2D Wannier exciton in a thin film that interacts with phonons via the longitudinal optical phonon-exciton Frohlich interaction. We show that the superradiant properties of the coherent state in hybrid perovskites are stable against perturbations caused by the longitudinal optical phonon-exciton Frohlich interaction. Using the multiconfiguration Hartree approach, we derive semiclassical equations of motion for a single-exciton wavefunction, where the vibrational degrees of freedom interact with the Wannier exciton through a mean-field Hartree term. Superradiance is effectively described by a non-Hermitian term in the Hamiltonian. The analysis was then extended to multiple excited states using the semiclassical Hamiltonian as the basic model.
We demonstrate that the ground state of the model exciton Hamiltonian with long-range interactions is a symmetric Dicke superradiant state, where the Frohlich interaction is nullified. The additional density matrix-based consideration draws an analogy between this system and stable systems, where the conservation laws determine the nullification of the constant (momentum-independent) decay rate part. In the exciton-phonon system, nullification is associated with the absence of a momentum-independent component in the Wannier exciton-phonon interaction coupling function.

\end{abstract}
\maketitle

\section{Introduction}\label{sec:description}
Macroscopic quantum phenomena are widely used in modern quantum technologies. One example is quantum computing based on  exciton-polariton (EP) condensates \cite{Ghosh2020npj, Kavokin24}. Quantum coherence is the main requirement for observing macroscopic quantum phenomena such as superconductivity, Bose-Einstein condensation, and superfluorescence (SF); therefore, except for the EP condensate, quantum coherence can be achieved under cryogenic conditions. Recently, SF in hybrid perovskite thin films was first observed by the North Carolina State University groups at a temperature of 78 K in methyl ammonium lead iodide (MAPbI$_{\text{3}}$ thin film) \cite{Gundogdu2021Nature_Phot}, and then in quasi-two-dimensional (2D) phenethylammonium caesium lead bromide (PEA:CsPbBr$_{\text{3}}$) at room temperature~\cite{Gundogdu2022Nature_Phot}. The discovery of high-temperature SF in hybrid perovskites naturally raises questions regarding the underlying mechanism. The physical picture of the high-temperature SF drawn from the analysis of the experiments conducted by Gundogdu et al.~\cite{Gundogdu2022Nature_Phot} suggests that the formation of large polarons protects the electronic excitation from dephasing even at room temperature. This study aims to elucidate the physical mechanism that enables SF in hybrid perovskites at high temperatures. One of our goals is to formulate a rigorous theoretical approach capable of describing the collective optical effects in systems of Wannier excitons that interact with vibrations. We utilize the multiconfiguration Hartree approach, which is similar to that used to describe collective phenomena in polariton systems~\cite{Osipov_Fainberg23PRB}, to study SF in Wannier exciton-phonon systems. The approach results in a set of coupled nonlinear differential equations of motion, which describe a system of interacting oscillators with collective behavior. The solution of nonlinear equations is a nontrivial problem by itself; therefore, to focus on the new physical phenomenon as a first step, we consider the model in a simplified formulation.

Many atoms or molecules interacting cooperatively with a shared optical field produce a collective optical response that differs from the response generated by independent atoms or molecules \cite{Dicke54PR}. Specifically, the SF phenomenon is caused by the fact that the collective radiation rate of multiple emitters exceeds the emission rate of their spontaneous luminescence. The simplest theoretical model for describing superradiance is based on the effective non-Hermitian Hamiltonian, originally suggested in nuclear theory, to describe the temporal propagation of an open quantum system~\cite{Auerbach_Zelevinsky11RPP, Fainberg2016CP, Nitzan22PRA}, where the Hermitian part describes the quantum system of $N$ identical two-level molecules. The non-Hermitian part of the effective non-Hermitian Hamiltonian represents the spontaneous emission. The superradiant state, which is a coherent superposition of all molecular excitations, decays at an enhanced superradiance rate, which is $N$ times larger than the single-molecule emission rate.  The more sophisticated Dicke model is characterized by a second-order parametric phase transition, known as the superradiant phase transition~\cite{Hepp_Lieb73, Larson_Irish2017}.  One distinguishes between the static and dynamic characteristics of the critical phenomena. The static characteristics include equilibrium quantities, such as susceptibility and magnetization. The dynamic critical phenomena are determined by the time evolution parameters of the system, such as the relaxation characteristic times. As the first step of our analysis, we focus on studying the dynamic critical phenomenon that is the decay of the superradiant state in the presence of vibrations. In particular, this allows us to restrict the model to collective emission from single excitation states~\cite{Scully10PRA, Nitzan22PRA}. We extend our analysis to the original Dicke superradiance problem \cite{Dicke54PR}, that is when a significant part of the molecules (excitons) is excited, utilizing the isotropic Lipkin--Meshkov--Glick (LMG) model \cite{LMG1_1965, LMG2_1965, LMG3_1965, Vidal2007, Carrasco2016}. In our consideration, we distinguish between the 3D and quasi-2D (thin film) set-ups. It is known (see, e.g., \cite{Knoester92PRL}) that the general dispersion relation connects the superradiant-exciton regime at film thicknesses smaller than the optical wavelength with bulk polaritons in thick crystals. This motivated us to search for the superradiant regime in quasi-2D crystals.

The remainder of this paper is organized as follows. In the next section, we define the Hamiltonian for a Wannier exciton interacting with longitudinal optical (LO) phonons in polar crystals (Section \ref{sec:SFinPolarCrystals}). We also derived the exciton-phonon interaction Hamiltonian for quasi-2D excitons and LO phonon modes. In Section \ref{sec:Equations} we derive equations describing vibration-assisted single-exciton wavefunction evolution within the mean-field Hartree approximation. In Section \ref{sec:small_deviations}, the equations of motion are solved for small deviations from the superradiant state. Hartree theory is used to obtain the semiclassical Hamiltonian, the exciton part of which is modelled by the isotropic LMG model (Section \ref{sec:SC_Hamiltonian}). The ground state of the exciton part of the semiclassical Hamiltonian is the symmetric Dicke superradiant state. In Section \ref{sec:DM}, we reformulate our theory in terms of the density matrix. Section \ref{sec:PT} provides a perturbative solution for our model, and its comparison with the solution for a model of $N$-identical two-level molecules. We demonstrate the stability of the superradiant state in hybrid perovskite thin films with respect to the dephasing caused by the LO phonon-exciton Frohlich interaction. Finally, in Section~\ref{sec:Conclusion} we provide a brief conclusion.

\section{The Model Hamiltonian}\label{sec:SFinPolarCrystals}

The Hamiltonian for a Wannier exciton interacting with phonons typically
consists of three components: the Wannier exciton energy operator ($\hat
{H}_{ex}^{W}$), phonon energy operator ($\hat{H}_{ph}$), and Wannier
exciton-phonon interaction energy operator ($\hat{H}_{I}^{W}$)
\begin{equation}
\hat{H}_{0}=\hat{H}_{ex}^{W}+\hat{H}_{ph}+\hat{H}_{I}^{W},\label{eq.:H_03}%
\end{equation}
where the free exciton energy is
\begin{equation}
\hat{H}_{ex}^{W}=\hbar\sum_{n\bm{k}}W_{n}(\bm{k})\hat{B}_{n\bm{k}}%
^{\dag}\hat{B}_{n\bm{k}}, \label{eq:Wex_hamilt4}%
\end{equation}
where the summation over $\bm k$ runs over the first Brillouin zone, $n$ is the principal quantum number of the excitonic state ($n=1,2,3,\dots$ for the 3D case and $n=0,1,2,\dots$ for 2D like that of the hydrogen atom). The free phonon part of the Hamiltonian $\hat{H}_0$ is
\begin{equation}
\hat{H}_{ph}=\sum_{\bm{q},s}\hbar\omega_{s}(\bm{q})\hat{b}_{\bm{q}s}^{\dag}\hat{b}_{\bm{q}s} \label{eq.:H_ph}
\end{equation}
We used the $s$ branch index in the phonon Hamiltonian because this form of the
interaction operator can describe both acoustic and optical phonons.
Typically, in the case of Wannier excitons in polar crystals one can focus on the
interactions with optical phonons alone. The operators $\hat{B}_{n\bm{k}}^{\dag}$ ($\hat{B}_{n\bm{k}}$) and $\hat{b}_{\bm{q}s}^{\dag}$ ($\hat{b}_{\bm{q}s}$) are the creation (annihilation) operators of excitons and
phonons, respectively. 

Hybrid perovskite thin films were used in SF experiments~\cite{Gundogdu2021Nature_Phot, Gundogdu2022Nature_Phot}. We previously noted that the superradiant exciton regime exists in thin films that are thinner than the optical wavelength, whereas bulk polaritons exist in thick crystals \cite{Liu_Lee80, Knoester92PRL}. There is an additional reason why quasi-2D crystals are well-suited for observing exciton luminescence. To observe exciton luminescence, the thermal energy $k_{B}T$ (where $k_{B}$ is Boltzmann's constant and $T$ is the temperature) must be less than the exciton ground state binding energy $E_{bind}$, $k_{B}T<E_{bind}$. In the 3D case, $E_{bind}$ coincides with the exciton Rydberg energy $E_{R}$, $E_{bind}=E_{R}$, whereas in 2D, $E_{bind}=4E_{R}$ \cite{Hau01}. At 330 K, the thermal energy $k_{B}T$ is approximately 28 meV. In other words, to observe exciton luminescence from quasi-2D crystals at room temperature, $E_{R}$ must exceed 7 meV. For comparison, Long et al. \cite{Long19Nanoscale} studied the exciton-phonon interaction of quasi-2D (PEA)$_{2}$(CsPbBr$_{3}$)$_{n-1}$PbBr$_{4}$ perovskite with varying CsPbBr$_{3}$ layer numbers $n$. They evaluated the binding energy of pure phase (PEA)$_{2}$PbBr$_{4}$ to be 163~meV. Their findings also demonstrated the amplification of the exciton-phonon interaction when the quantum-well structure became thinner. Large polarons are formed in the halide perovskites \cite{Miyata17}. Therefore, this enhanced exciton-phonon interaction can be attributed to the LO phonon-exciton Frohlich interaction. Thus, we consider a
polaron, where an electron interacts with the LO phonons of frequency $\omega_{l}$, under the assumption that the phonon dispersion can be neglected, that is $\omega_{l}(\mathbf{k})=\omega_{l}$ \cite{Frohlich54,Feynman55PR,Haken56,Kit63,Feynman98,Jai_Singh94}. Then the free phonon Hamiltonian, Eq.(\ref{eq.:H_ph}), becomes $\hat{H}_{ph}=\hbar\omega_{l}\sum_{\bm{q}}\hat{b}_{\bm{q}}^{\dag}\hat{b}_{\bm{q}}$.

Within the quasi-2D perovskite structure, 3D domains and layered 2D domains with varying thicknesses coexist in the thin film, as discussed in the Supplementary Information of Ref.\cite{Gundogdu2022Nature_Phot}. Interestingly, the spectral properties of the 2D domains with three or more
layers are minimally affected by quantum confinement. Consequently, we aim to deduce the interaction Hamiltonian between quasi-2D excitons and lattice phonon modes.
To formulate the interaction Hamiltonian we follow the method introduced in Refs.~\cite{Haken56,Pollmann77}. Consider an excited electron in the conduction band  and a hole located in the valence band; their motion is correlated due to the influence of one on the other. The 3D interaction Hamiltonian for the Wannier exciton-phonon in the center-of-mass representation is
\begin{equation}
\hat{H}_{I}^{W}=-i\hbar\sum_{\bm{k}}F_{\bm{k}}(\bm{r})\exp
(i\bm{k}\cdot\bm R)(\hat{b}_{\bm{k}}-\hat{b}_{-\bm{k}}^{\dag}),\label{eq.:H^pol(W)_Ic}
\end{equation}
with the amplitude
\begin{equation}
F_{\bm{k}}(\bm{r})=\omega_{l}\sqrt{\frac{4\pi\alpha}{uV}}
\frac{1}{k}\left[  e^{ip_{e}\bm{k}\cdot \bm r}  - e^{-ip_{h}\bm{k}\cdot
\bm r}\right],  \label{eq.:F_k(r)}
\end{equation}
where $\alpha_{e}=\frac{1}{2}\left(\frac{1}{\varepsilon_{\infty}}-\frac{1}{\varepsilon_{0}}\right)  \frac{e^{2}}{\hbar\omega_{l}}u_{e}$ is the
Frohlich coupling constant~\cite{Frohlich54,Feynman98,Jai_Singh94,Miyata17}, $e$ is the carrier charge, $\varepsilon_{\infty}$ and $\varepsilon_{0}$ are
optical and static dielectric constants, $u_{e}=\sqrt{2m_{e}^*\omega_{l}/\hbar}$, $\bm{r}=\bm{r}_{e}-\bm{r}_{h}$ is the relative coordinate, and $\bm{R}$ is the exciton center-of-mass coordinate, $\bm{R}=(m_{e}^*\bm{r}_{e}+m_{h}^*\bm{r}_{h})/M^*$, $m_{e}^*$ and $m_{h}^*$ are the electron and hole effective masses, respectively; $M^*=m_{e}^*+m_{h}^*$. The coefficients $p_{i}\equiv m_{r}/m_{i}^*$, $i=e,h$; $1/m_{r}=1/m_{e}^*+1/m_{h}^*$ is the inverse reduced mass, and $V$ is the quantization volume for the phonon field. Evidently, that $\alpha/u$ does not depend on the effective particle mass. Therefore, we
omitted the particle index in Eq.(\ref{eq.:F_k(r)}).

The exciton state with a given exciton center-of-mass momentum $\bm{q}'$ is defined by the wave function $\ket{n\bm{q}'} \equiv \ket{ \Phi_{n,\bm{q}'}(\bm{R},\bm{r})} =\exp(i\bm{q}'\cdot \bm{R})\psi_{n}(\bm{r})/\sqrt{V_{0}}$ in the 3D case where $\psi_{n}(\bm{r})$ is the normalized hydrogen atom wave function, $n$ is the excitation level number, and $V_{0}$ is the crystal volume. The eigenenergy of the eigenfunction $\Phi_{n,\bm{k}}(\bm{R},\bm{r})$ is known to be $\hbar W_{n}(\bm{k})=E_{c}(\bm{k}_{0})-E_{v}(\bm{k}_{0})+E_{n}+\frac{\hbar^{2}k^{2}}{2M^*}$, where $E_{n}$ is the energy of the internal structure of an exciton in a state $n$, $E_{c}(\bm{k}_{0})-E_{v}(\bm{k}_{0})$ is the energy gap between the minimum of the first conduction band and the maximum of the valence band.

Next we introduce the basis set for the single-exciton states and exciton operators. The basis set consists of the vectors $|n\bm{q}\rangle$ obtained by acting on the vacuum vector $|0\rangle$ (no excitons), which describes the ground state of the exciton subsystem, with the exciton creation operator $\hat{B}_{n\bm{q}}^{\dag}=\ket{n\bm{q}}\bra{0}$ accompanying by the excitation of state $n$. That is, $\ket{n\bm{q}}=\hat{B}_{n\bm{q}}^{\dag}\ket{0}$.

Excitons in a thin film maintain the dual merits of a coherent nature on the two-dimensional plane of the film (referred to as a quantum well) and
superradiant decay in one direction perpendicular to the thin film. We will now examine a system that satisfies the following condition for its thickness,
denoted as $l$: $l\sim a_{0}<<\lambda\leq L$. Here, $a_{0}$ represents the exciton Bohr radius, $\lambda$ denotes the wavelength of light, and $L^{2}$
corresponds to the area of the thin film. To analyze this system, we decompose the wave vector $\bm{k}$ into two components: $k_{z}\hat{z}$ along the
perpendicular direction ($z$-axis) and $\bm{q}$ parallel to the surface. Exciton state $n$ with the lowest energy can be described by
(compare with Eq.(8) of Ref. \cite{Hanamura88} with corrected misprints)
\begin{multline}
 \ket{n\bm{q}}=\Psi_{\bm{q}n,1,1}(\bm{R},\bm{r},z_{e},z_{h})\\
=\frac{1}{\sqrt{L^{2}}}\exp(i\bm{q}\bm{R})\psi_{n}
(\bm{r})f_{1}(z_{e})f_{1}(z_{h}) \label{eq:2Dwavefunction}
\end{multline}
where $f_{m}(z)=\sqrt{2/l}\sin\left(  m\pi z/l\right)$. The two-dimensional vectors $\bm{R}$ and $\bm{r}$ represent the exciton coordinates in the film plane, describing the center-of-mass motion and the electron-hole relative motion, respectively.

Expansion of the coordinate dependent amplitude in Eq.~(\ref{eq.:H^pol(W)_Ic}) over the exciton basis states can be done with the help of the identity operator $\sum_{n\bm{q}}\ket{n\bm{q}}\bra{n\bm{q}}=1$ to obtain
\begin{multline}
F_{\bm{k}}(\bm{r})\exp(i\bm{k}\cdot \bm R)  \\  =\sum_{nn'\bm{q}\bm q'}
\ket{n\bm{q}}
\bra{n\bm{q}}F_{\bm{k}}(\bm{r})\exp(i\bm{k}\cdot \bm R)\ket{
n'\bm{q}'} \bra{n'\bm{q}'}
\\
 =\sum_{nn'}\langle n|F_{\bm{k}}(\bm{r})\left\vert
n'\right\rangle \sum_{\bm{qq}'}\langle\bm{q}%
|\exp(i\bm{k\cdot R})\left\vert \bm{q}'\right\rangle
\hat{B}_{n\bm{q}}^{\dag}\hat{B}_{n'\bm{q}'}.
\end{multline}
The matrix element $\bra{\bm{q}}\exp(i\bm{k}\cdot \bm R)\ket{\bm{q}'}=\delta_{\bm{q}'\bm{-q}+\bm{k},\bm{0}}$, then  exciton-phonon interaction operator $\hat{H}_{I}^{W}$, Eq.(\ref{eq.:H^pol(W)_Ic}), can be expressed in a form that
contains both exciton and phonon operators:
\begin{equation}
\hat{H}_{I}^{W}=-i\hbar\sum_{nn'\bm{k}\bm q}D_{l}^{pol}(\bm{k};nn')\hat{B}_{n,\bm{q}+\bm{k}}^{\dag}\hat{B}_{n'\bm{q}}(\hat{b}_{\bm{k}}-\hat{b}_{-\bm{k}}^{\dag}), \label{eq.:H^pol(W)_Id}
\end{equation}
where the electronic matrix element for the coupling function of the Wannier exciton-phonon interaction is given by
\begin{equation}
D_{l}^{pol}(\bm{k};nn')=\omega_{l}\sqrt{\frac{4\pi\alpha}{uV}}\frac{1}{k}[Q_{e}(\bm{k};nn')-Q_{h}(\bm{k};nn')]
\label{eq.:D^pol_l(k,r)}%
\end{equation}
The hermiticity of the interaction Hamiltonian, $\hat{H}_{I}^{W}=\hat{H}_{I}^{W\dag}$, requires the function $D_{l}^{pol}(\bm{k};nn')$ to
satisfy a certain symmetry: $D_{l}^{pol}(\bm{k};nn')=D_{l}^{\ast pol}(-\bm{k};n'n)$. The function $Q_{e(h)}(\bm{k};nn')\equiv\int d^3\bm{r}\psi_{n}^*(\bm{r})\psi_{n'}(\bm{r})\exp(\pm i p_{e(h)}\bm{k}\bm r)$ is the Fourier transform of the electron (hole) charge distributions in the internal exciton motion~\cite{Toyozawa58, Toyozawa59}. It defines the effectiveness of the electron (hole) interaction with a particular phonon $\bm{k}$. Because the wave functions $\psi_{n}(\bm{r})$ and $\psi_{n'}(\bm{r})$ are orthonormal, the function $Q_{e(h)}(\bm{k};nn')$ tends to unity at $n=n'$ and to zero for $n\neq n'$ as $\bm{k}$ tends to zero. The function $Q_{e(h)}(\bm{k};nn')$ for any combination of $n$ and $n'$ becomes very small when $2\pi/k<a_{0}$, that is when the phonon wavelength $2\pi/k$ is smaller than the mean distance between the electron and the hole --
the exciton Bohr radius $a_{0}$.

\subsection{The interaction amplitude $D_{l}^{pol}(\mathbf{k};nn')$ in the quasi-2D model}
The quantity $D_{l}^{pol}(\bm{k};nn')$, Eq.(\ref{eq.:D^pol_l(k,r)}), can be calculated analytically for the quasi-2D model.
For exciton state $n=n'=0$, the integral on the right-hand side of Eq.(\ref{eq.:D^pol_l(k,r)}), can be written as (without loss of generality the $x$ axis can be chosen to be collinear with the $\bm{k}$ vector):
\begin{equation}
\frac{8}{\pi a_{0}^{2}} \int_{0}^{\infty}r dr e^{-\frac{4r}{a_{0}}} \int\limits_{0}^{2\pi} d\varphi\left[e^{ip_{e}kr\cos\varphi}-e^{-ip_{h}kr\cos\varphi}\right],
\end{equation}
where we used the ground-state wave function $\psi_{0}(\bm{r})=\sqrt{\frac{8}{\pi a_{0}^{2}}}\exp\left(-\frac{2r}{a_{0}}\right)$. The integral over $\varphi$ of $\exp(iz\cos\varphi)$ is the integral representation of the Bessel function $J_{0}(z)$ multiplied by $2\pi$, and the integration over $r$ can be done using the table formulas~\cite{Prudnikov98}:
\begin{multline}
D_{l}^{pol}(\bm{k};00)=a_0\omega_{l}\sqrt{\frac{4\pi\alpha}{uV}}\;\frac{1}{a_0k}\\\times
\left[  \left(  1+\left(\frac{p_{e}ka_{0}}{4}\right)^{2}\right)^{-
3/2}-\left(  1+\left(\frac{p_{h}ka_{0}}{4}\right)^{2}\right)^{-3/2}\right]
\label{eq.:D^pol_l(k)2D}
\end{multline}

For the small $k$ values, the function $D_{l}^{pol}(\bm{k};00)$ is approximated by
\begin{multline}
D_{l}^{pol}(\bm{k};00)=a_{0}\omega_{l}\sqrt{\frac
{4\pi\alpha}{uV}}\Big[(p_{h}^{2}-p_{e}^{2})\;\frac{3 a_{0}k}{32} \\
+ (p_{h}^{4}-p_{e}^{4})\frac{15 a_0^3 k^3}{2048}+\mathcal{O}(k^5) \Big]\label{eq.:D^pol_l(k->0)2D}
\end{multline}
In other words, the interaction coefficient $D_{l}^{pol}(\bm{k};00)$ approaches zero when $k$ approaches $0$. In contrast, the Frohlich interaction of LO phonons with electrons and holes does not go to zero as $\bm{k}$ approaches $0$~\cite{Frohlich54,Feynman55PR,Haken56,Kit63,Feynman98,Jai_Singh94}. Physically, this can be explained by the fact that exciton is confined in space, while the quantum electron wave is stretched over the whole crystal as well as the zero $k$ phonon wave. Note that, despite the fact that Eq.(\ref{eq.:D^pol_l(k)2D}) for the  Wannier exciton-phonon interaction coupling function $D_{l}^{pol}(\bm{k};00)$ differs from that derived for bulk crystals, $D_{l}^{pol}(\bm{k};1s1s)$, in Refs.~\cite{Toyozawa59, Cardona2010}, the proportionality of $D_{l}^{pol}(\bm{k})$ to the wave vector at a small $k$ remains in both cases.

The non-diagonal term $D_{l}^{pol}(\bm{k};0(1,0))$ can be calculated in a similar manner. Considering that the wave function $\psi_{1,0}(\bm{r})$ is given by $\psi_{1,0}(\bm{r})=\sqrt{\frac{8}{27\pi a_{0}^{2}}}\exp\left( -\frac{2r}{3a_{0}}\right)  (1-\frac{4r}{3a_{0}})$, the resulting expression is
\begin{multline}
D_{l}^{pol}(\bm{k};0(1,0))=a_{0}\omega_{l}\sqrt{\frac{4\pi\alpha}{uV}}
\frac{27\sqrt{3}}{512}\; a_{0}k\\\times \left[\frac{p_{e}^{2}}{\left[\left(\frac{3a_{0}
p_{e}k}{8}\right)^{2}+1\right]^{5/2}}-\frac{p_{h}^{2}}{\left[\left(\frac{3a_{0}p_{h}k}{8}\right)^{2}+1\right]^{5/2}}\right]\label{eq.:D^pol_l(k;01)2D_2}
\end{multline}

\begin{figure}
\begin{center}
\begin{tabular}{lc}
a) &\includegraphics[scale=0.35]{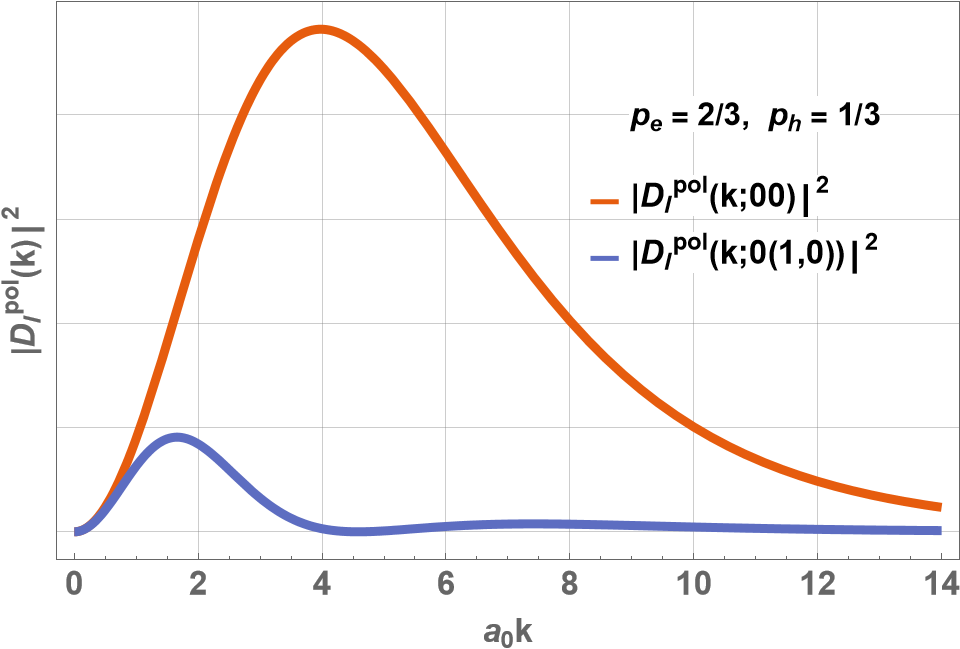}\\
b)&\includegraphics[scale=0.35]{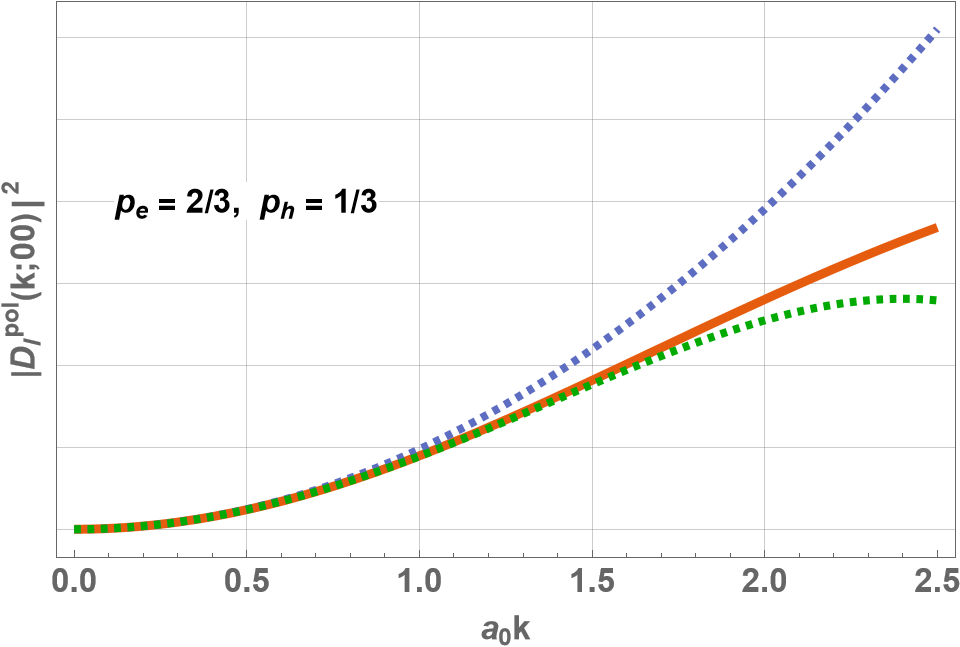}
\end{tabular}
\caption{\label{fig:D2} a) The functions $\abs{D_{l}^{pol}(\bm{k};00)}^2$ (Eq.~\ref{eq.:D^pol_l(k)2D}) and $\abs{D_{l}^{pol}(\bm k;0(0,1))}^2$ (Eq.~\ref{eq.:D^pol_l(k;01)2D_2}) in units $a_0\omega_{l}\sqrt{\frac{4\pi\alpha}{uV}}$ plotted v.s. the wavevector $a_0 k$.  b) The function $|D_{l}^{pol}(\bm{k};00)|^{2}$ (orange) plotted in the vicinity of small $k$ and its power series expansions: (blue) up to the quadratic term $(a_0k)^{2}$ and (green) up to the fourth power $(a_0 k)^{4}$ (Eq.~\ref{eq.:D^pol_l(k->0)2D}). For both plots $p_{e}=2/3$ and  $p_{h}=1/3$.
\label{fig:D2_approximate}}
\end{center}
\end{figure}

Graphs of the functions $|D_{l}^{pol}(\bm{k};00)|^{2}$ and $|D_{l}^{pol}(\bm{k};0(1,0))|^{2}$  are shown in Fig.\ref{fig:D2} for a particular choice $p_{e}=2/3$ and $p_{h}=1/3$. One can see that for small $k$ values, the behavior of $D_{l}^{pol}(\bm{k};0(1,0))\sim k(p_{e}^{2}-p_{h}^{2})$ is similar to that of the diagonal term $D_{l}^{pol}(\bm{k};00)$, Eq.(\ref{eq.:D^pol_l(k->0)2D}). Both plots are equal to zero at zero $k$. For a small non-zero $k$, the denominators in Eq.(\ref{eq.:D^pol_l(k)2D}) can be expanded in $k$ to the first order, yielding the first term on the right-hand side of Eq.(\ref{eq.:D^pol_l(k->0)2D}). Both functions $|D_{l}^{pol}(\bm{k};00)|^{2}$ and $|D_{l}^{pol}(\bm{k};0(1,0))|^{2}$ increase as $k^2$ for $ka_{0}<<1$. The function $|D_{l}^{pol}(\bm{k};00)|^{2}$
reaches its maximum at $ka_{0}\approx 4$, then drops to zero as $k^{-8}$ for large $ka_{0}$. The function $|D_{l}^{pol}(\bm{k};0(1,0))|^{2}$
maximum is reached at $ka_{0} \approx 1.5$ and then also decreases to zero for large $ka_{0}$. The maximum value of the non-diagonal term $|D_{l}
^{pol}(\bm{k};0(1,0))|^{2}$ is significantly less than that of the diagonal term $|D_{l}^{pol}(\bm{k};00)|^{2}$. Notably, both terms, Eq.~(\ref{eq.:D^pol_l(k)2D}) and Eq.~(\ref{eq.:D^pol_l(k;01)2D_2}), are equal to zero exactly for all $k$ when the electron and hole masses are equal $m_{e}^*=m_{h}^*$. This can be explained by the fact that equal electron and hole masses mean that the center of mass and the center of charge distribution coincide, so that the electric fields affecting the electron and hole cancel out.

\subsection{The superradiance effective Hamiltonian}
As noted in the Introduction, the simplest theoretical model for describing superradiance is based on the effective non-Hermitian Hamiltonian $\hat{H}_{eff}=\hat{H}_{M}-i\hbar\frac{\Gamma}{2}\hat{P}$~\cite{Auerbach_Zelevinsky11RPP, Fainberg2016CP, Nitzan22PRA}, where the Hermitian Hamiltonian $\hat{H}_{M}$ describes the quantum system of $N$ identical two-level molecules. The non-Hermitian operator $-i\frac{\Gamma}{2}\hat{P}$ represents spontaneous emission with a single molecule emission rate $\Gamma$ and a projection operator $\hat{P}$. The operator $\hat{P}=\sum_{m,n=1}^{N} \ket{me}\bra{ne}$ uniformly distributes excitations over all two-level molecules, where $\ket{me}$ stands for the excited state of the $m$-th molecule, while all others are in the ground state. In the absence of the intermolecular Coulomb interactions, the effective Hamiltonian has one complex-valued eigenvalue corresponding to the superradiant state $\ket{SR}=\frac{1}{\sqrt{N}}\sum\limits_{m=1}^{N}\ket{me}$, which is a coherent superposition of all molecular excitations. $|SR\rangle$ decays at an enhanced superradiance rate $N\Gamma$. The remaining $N-1$ real eigenvalues are referred to as non-decaying dark states. Following these arguments, below (in Section III), we modify the Hamiltonian of our system by introducing a non-Hermitian part.

\section{Equations of motion for a single-exciton wavefunction}
\label{sec:Equations}
This section aims to develop a rigorous approach based on fundamental principles for describing vibration-assisted single-exciton wavefunction evolution. As mentioned earlier, our initial focus is on models that involve collective emission by single excitation states~\cite{Scully10PRA, Nitzan22PRA}. The Hartree approximation was used to derive these equations. The derivation is similar to that presented in Ref.\cite{Osipov_Fainberg23PRB}, which describes the evolution of the polariton-vibration wavefunction.

To construct the single-exciton wavefunction, we expand it on the previously introduced basis set of the single-exciton states $\ket{n\bm{q}}
=\hat{B}_{n\bm{q}}^{\dag}\ket{0}$ (see Eq.~\ref{eq:2Dwavefunction}). Therefore, the exciton wavefunction is defined by the sum $\sum_{n\bm{q}}C_{n}(\bm{q}|t)\ket{n\bm{q}}$, where the time-dependent coefficients $C_{n}(\bm{q}|t)$ represent the exciton wavefunction in the single-exciton basis.


We use the coherent state basis $\ket{\sigma}$~\cite{Glauber63, BBGK1971} as the basis for the phonon states. Each coherent state $|\sigma\rangle$ is parametrized by a multidimensional complex-valued vector $\sigma$ that encodes the coherent state center, i.e. the classical coordinate $x$ and the classical momentum $p$, namely $\sigma=x+ip$. The vibration operators act on the basis vectors as: $b\ket{\sigma}=\sigma\ket{\sigma}$, $\bra{\sigma}b^\dag
=\bra{\sigma}\sigma^*$. In this regard, it is worth mentioning the
complex-valued classical coordinates that may be interpreted as 
eigenvalues associated with the coherent state of the harmonic oscillator
\cite{Miyazaki24}. The normalized coherent state has the following representation
\begin{equation}
\ket{\sigma} = e^{-\frac{1}{2}\abs{\sigma}^2}e^{\sigma b^\dag}\ket{0}, \label{cohState}
\end{equation}
where the state $\ket{0}$ is the ground state of the corresponding oscillator. We now follow the standard scheme used in the multiconfiguration Hartree approach~\cite{MMC1992, WG2019, Osipov_Fainberg23PRB}. The advantage of this approach is that the resulting dynamic equations are much simpler for analysis than the original Schr\"odinger equation; the equations are similar to those of the mean-field Hartree theory. To describe the wavefunction time evolution $|\Psi(t)\rangle$, we employ the time-dependent basis vectors Ansatz (see also Davydov Ansatz~\cite{Davydov1973,Gelin2015} used in the theory of one-dimensional molecular aggregates). It assumes that the vibration part of the basis vectors $\ket{\bm \sigma(t),n\bm q}$ depend on time in addition to the time-dependent expansion coefficients $C_{n}(\bm{\sigma},\bm{q}|t)$. Thus, our working basis consists of the direct products of the exciton and
vibrational states:
\begin{equation}
\ket{\bm{\sigma},n\bm{q}}=\ket{\bm{\sigma}}\ket{n\bm{q}},\quad\ket{\bm{\sigma}}=\bigotimes_{\bm{k}}\ket{\sigma
_{\bm{k}}},\label{ketsigmaa}
\end{equation}
With this, the wavefunction is expanded as follows
\begin{equation}
|\Psi(t)\rangle=\sum_{n\bm{q}}C_{n}(\bm{\sigma
},\bm{q}|t)|\bm{\sigma}(t),n\bm{q}\rangle\label{wavefunction1}%
\end{equation}

To proceed, we use the Dirac-Frenkel variation principle. This approach is known to be useful for describing quantum dynamics in systems with a large number of vibrational degrees of freedom~\cite{SC2004, AR2017, Miller2002, WG2019}. Wavefunction variation allows us to separate the time evolution of the vibrational subsystem from the quantum evolution of the exciton wavefunction. The derivation of the equations of motion follows a procedure similar to that used in our recent study on molecular polaritons~\cite{Osipov_Fainberg23PRB}. The variation of the wavefunction,
Eq.(\ref{wavefunction1}), is (here and below out of brevity we omit the time variable $t$ in the coefficients):
\begin{multline}\label{WFVar}
\delta\bra{\Psi}=\sum_{n\bm{q}}\bra{\bm{\sigma
},n\bm{q}}\Bigg\{\delta C_{n}^*(\bm{\sigma},\bm{q})\\
+C_{n}^{\ast
}(\bm{\sigma},\bm{q})\sum_{\bm{k}}\left[\delta\sigma_{\bm{k}}^*\hat{b}_{\bm{k}}-\frac{1}{2}(\sigma_{\bm{k}}\delta\sigma_{\bm{k}}^*+\sigma_{\bm{k}}^*\delta\sigma_{\bm{k}})\right]
\Bigg\}
\end{multline}
The second term on the right side of Eq.~(\ref{WFVar}) results from varying the coherent state expressed in the form of Eq.~(\ref{cohState}). From Eq.~(\ref{WFVar}) we also derive the expression for the wavefunction time derivative:
\begin{multline}
\frac{d}{dt}\ket{\Psi}=\sum_{n\bm{q}}\Bigg\{\dot{C}_{n}(\bm{\sigma},\bm{q})\\
+C_{n}(\bm{\sigma},\bm{q})\sum_{\bm{k}}\left[\dot{\sigma}_{\bm{k}}\hat{b}_{\bm{k}}^{\dag}-\frac
{1}{2}(\sigma_{\bm{k}}\dot{\sigma}_{\bm{k}}^*+\sigma_{\bm{k}}^*\dot{\sigma}_{\bm{k}})\right]\Bigg\}\ket{\bm{\sigma},n\bm{q}}\label{psivar}
\end{multline}
The dot over a variable, as the standard, represents the time derivative. By varying the Schr\"{o}dinger equation with Hamiltonian $\hat{H}_{0}$ (Eq.~\ref{eq.:H_03}), $\bra{\Psi(t)}i\hbar\frac{d}{dt} -\hat{H}_{0}\ket{\Psi(t)}=0$, with respect to the bra-vector and setting each term proportional to the independent variations $\delta C_{n}(\bm{\sigma}',\bm{q}',|t)$ and $\delta\sigma_{\bm{k}}$ to zero, yields the system of coupled equations:
\begin{eqnarray}
\bra{\bm{\sigma}',n'\bm{q}'}i\hbar\frac{d}{dt}-\hat{H}_{0}\ket{\Psi}& =&0\label{varH1}\\
\sum_{n'\bm{q}'}C_{n'}^*(\bm{\sigma}',\bm{q}')\bra{\bm{\sigma}',\bm{q}'}\hat{b}_{\bm{k}}\left(i\hbar\frac{d}
{dt}-\hat{H}_{0}\right) \ket{\Psi}&  =&0 \label{varH2}
\end{eqnarray}
Substituting Eq.~(\ref{psivar}) into Eqs.~(\ref{varH1}) and~(\ref{varH2}) yields the following two equations:
\begin{multline}\label{eq1}
\dot{C}_{n'}(\bm{\sigma},\bm{q}')
=\frac{1}{2} C_{n'}(\bm{\sigma},\bm{q}')\sum_{\bm{k}}(\sigma_{\bm{k}}\dot{\sigma}_{\bm{k}}^*-\sigma_{\bm{k}}^*\dot{\sigma}_{\bm{k}})\\
-\frac{i }{\hbar}\sum_{n\bm{q}}\frac{\braket{\bm{\sigma}',n'\bm{q}'|\hat{H}_{0}|\bm{\sigma},n\bm{q}}}{\braket{\bm{\sigma}'|\bm{\sigma}}} C_{n}(\bm{\sigma},\bm{q})
\\-C_{n'}(\bm{\sigma},\bm{q}')\sum_{\bm{k}}({\sigma'}_{\bm{k}}^*-\sigma_{\bm{k}}^*)\dot{\sigma}_{\bm{k}}
\end{multline}
\begin{multline}\label{eq2}
C_{n'}^*(\bm{\sigma}',\bm{q}') \left[\sigma_{\bm{k}}\dot{C}_{n'}(\bm{\sigma},\bm{q}')+\dot{\sigma}_{\bm{k}}C_{n'}(\bm{\sigma},\bm{q}')\right] \\
 =-\frac{i}{\hbar}\sum_{n\bm{q}}\frac{\braket{\bm{\sigma}',n'\bm{q}'|\hat{b}_{\bm{k}}\hat{H}_{0}|\bm{\sigma},n\bm{q}}}{\braket{\bm{\sigma}'|\bm{\sigma}}} C_{n'}^*(\bm{\sigma}',\bm{q}')C_{n}(\bm{\sigma},\bm{q}) \\
+\frac{1}{2}C_{n'}^*(\bm{\sigma}',\bm{q}')C_{n'}(\bm{\sigma},\bm{q}')
\\\times
\sigma_{\bm{k}}\sum_{\bm{k}'}[(\sigma_{\bm{k}'}\dot{\sigma}_{\bm{k}'}^*-\sigma_{\bm{k}'}^*\dot{\sigma}_{\bm{k}'})-2\dot{\sigma}_{\bm{k}'}\left(
{\sigma'}_{\bm{k}'}^*-\sigma_{\bm{k}'}^*\right)  ]
\end{multline}
The terms containing the Hamiltonian $\hat{H}_{0}$ (terms $\braket{\bm{\sigma}',n'\bm{q}'|\hat{H}_{0}|\bm{\sigma},n\bm{q}}$ and $\braket{\bm{\sigma}',n'\bm{q}'|\hat{b}_{\bm k}\hat{H}_{0}|\bm{\sigma},n\bm{q}}$) are calculated in Appendix~\ref{AppA}. Substituting the results of Eqs.(\ref{eq:<q'|H_ex|q>}),
(\ref{eq:<q'|H_ph|q>}), (\ref{eq:<q'|b_ksH_ex|q>}), (\ref{eq:b_ksH_ph}), (\ref{eq:<q'|H_I|q>}) and (\ref{eq:<q'|b_ksH_I|q>}) from Appendix A into Eqs.~(\ref{eq1}) and~(\ref{eq2}), we obtain a complete set of equations of motion for the wavefunction of the system. The resulting equations have the following structural form for each $n$ and $\bm q$
\begin{equation}
 \bar{S}_{n}(\bm{\sigma},\bm{q})+ \sum_{\bm{k}'}({\sigma_{\bm{k}'}^{\prime\ast}-}\sigma_{\bm{k}'}^*)S_{n}(\bm{\sigma},\bm{k}',\bm{q})  =0 \label{eq1struc}
\end{equation}
and for each $n$ and for each $\bm k$ and $\bm q$
\begin{multline}
 C_{n}^*(\bm{\sigma}',\bm{q})\Bigg[S_{n}(\bm{\sigma},\bm{k},\bm{q})+\sigma_{\bm{k}}\bar{S}_{n}(\bm{\sigma},\bm{q}) \\
+\sigma_{\bm{k}}\sum_{\bm{k}'} \left( {\sigma'}_{\bm{k}'}^*-\sigma_{\bm{k}'}^*\right) S_{n}(\bm{\sigma},\bm{k}',\bm{q})\Bigg]=0, \label{eq2struc}
\end{multline}
where Eq.~(\ref{eq1struc}) and Eq.~(\ref{eq2struc}) follow from Eq.~(\ref{eq1}) and Eq.~(\ref{eq2}), respectively. The functionals $S_n(\bm{\sigma},\bm{k},\bm{q})$ and $\bar{S}_n(\bm{\sigma},\bm{q})$ represent certain combinations of the expansion coefficients $C_{n}(\bm{\sigma},\bm{q})$, functions $\sigma_{\bm k}$ and their time-derivatives. The explicit form of these functionals was used to derive Eq.~(\ref{Cdot1}), when we set $\bar{S}_n(\bm{\sigma},\bm{q})=0$ and Eq.~(\ref{eq:dsigma_ks/dt}) for the choice $S_n(\bm{\sigma},\bm{k},\bm{q})=0$. This choice of constraints is explained in the following paragraph.

A system of equations (Eqs.~\ref{eq1struc},~\ref{eq2struc}), where the functionals $S_{n}(\bm{\sigma},\bm{k},\bm{q})$ and $\bar{S}_{n}(\bm{\sigma},\bm{q})$ are considered as variables, must be satisfied for any arbitrary values of $\bm \sigma'$ and $\bm \sigma$. From this we conclude that the system has only trivial solutions $S_{n}(\bm{\sigma},\bm{k},\bm{q})=0$, and $\bar{S}_{n}(\bm{\sigma},\bm{q})=0$. This allows the formulation of the final motion equations. From the constraint $\bar{S}_{n}(\bm{\sigma},\bm{q})=0$ we obtain the differential equation for coefficients $C_{n}(\bm{\sigma},\bm{q})$, it is
\begin{multline}
\dot{C}_{n'}(\bm{\sigma},\bm{q}')=-iW_{n'}(\bm{q}')C_{n'}(\bm{\sigma},\bm{q}')
\\
-i\omega_{l}C_{n'}(\bm{\sigma},\bm{q}') \sum_{\bm{k}'}\abs{\sigma_{\bm{k}'}}^2
+\sum_{n\bm{k}'}\tilde{\alpha}_{l}(\bm{k}';nn') C_{n}(\bm{\sigma},\bm{q}'+\bm{k}')
\\
+\frac{1}{2}C_{n'}(\bm{\sigma},\bm{q}')\sum_{\bm{k}}(\sigma_{\bm{k}}\dot{\sigma}_{\bm{k}}^*-\sigma_{\bm{k}
}^*\dot{\sigma}_{\bm{k}}), \label{Cdot1}
\end{multline}
where we have introduced the electron-vibrational coupling parameter
\begin{equation}
\tilde{\alpha}_{l}(\bm{k};nn')=D_{l}^{pol\ast}(\bm{k};nn')(\sigma_{\bm{k}}^*-\sigma_{\bm{-k}}).
\label{eq.:alpha_s(k)}
\end{equation}
The reflection symmetry of the coupling function $D_{l}^{pol}(\bm{q};nn')$ ($D_{l}^{\ast pol}(\bm{q};nn')=D_{l}^{pol}(-\bm{q};n'n)$) means that the function $\tilde{\alpha}_{l}(\bm{k};nn')$ is anti-symmetric, namely $\tilde{\alpha}^*_{l}(\bm{k};nn')=-\tilde{\alpha}_{l}(-\bm{k};n'n)$.

The first simple sequence of Eq.~(\ref{Cdot1}) is the equation for the probability densities $\abs{C_{n'}(\bm{\sigma},\bm{q})}^2$, it is
\begin{multline}
\frac{d}{dt}\abs{ C_{n}(\bm{\sigma},\bm{q})}^2=\sum_{n'\bm{k}'}\tilde{\alpha}_{l}(\bm{k}';n'n)C_{n}^*(\bm{\sigma},\bm{q})C_{n'}(\bm{\sigma},\bm{q}+\bm{k}')\\
   +\sum_{n'\bm{k}'}\tilde{\alpha}^*_{l}(\bm{k}';n'n)C_{n}(\bm{\sigma},\bm{q})C_{n'}^*(\bm{\sigma},\bm{q}+\bm{k}'),
\label{eq:d|C(q'|t)|^2/dt}
\end{multline}
which after summation over all $\bm q$ and $n$ reduces to the equation with the trivial right-hand side owing to the anti-symmetry of the electron-vibrational coupling parameter $\tilde{\alpha}_{l}(\bm{k};nn')$, that is
\begin{equation}
\frac{d}{dt}\sum_{n'\bm{q}'}\abs{C_{n'}(\bm{\sigma},\bm{q}'|t)}^2=0
\label{eq:conservation1}
\end{equation}
In other words, the exciton-phonon interactions do not change the number of excitations.

We now formulate the equations of motion for our model. The first equation of motion, for the vibration degrees of freedom, is obtained from the equation  $S_{n}(\bm{\sigma},\bm{k},\bm{q})=0$, which is written explicitly as follows:
\begin{equation}
\dot{\sigma}_{\bm{k}}=-(i\omega_{l}+\gamma)\sigma_{\bm{k}}+Tr[\hat{D}_{l}
^{pol}(-\bm{k})\hat{F}(\bm{k}|t)] \label{eq:dsigma_ks/dt}
\end{equation}
where we introduced a small decay $\gamma$ of the mode $\omega_{l}$, and the mean-field Hartree term is the quantum average of the exciton field $Tr[\hat{D}_{l}^{pol}(-\bm{k})\hat{F}(\bm{k}|t)]$. The trace operation denotes the summation over the exciton internal states $\sum_{nn'}D_{l}^{pol}(-\bm{k};n'n)F_{nn'}(\bm{k}|t)$ taking with the diagonal entries of the exciton density matrix, namely the matrix $F_{nn'}(\bm{k}|t)=\sum_{\bm{q}}\rho_{n,\bm{q}+\bm{k};n'\bm{q}}(t)$, where the density matrix of the single exciton state is defined as
\begin{equation}
\rho_{n,\bm{q}+\bm{k;}n'\bm{q}}(t)=\frac{C_{n'}^*(\bm{\sigma},\bm{q})C_{n}(\bm{\sigma},\bm{q}+\bm{k})}{\sum_{n'\bm{q}'}\abs{C_{n'}(\bm{\sigma},\bm{q}')}^2} \label{eq:DM}
\end{equation}
Clearly, that the matrix entries $F_{nn'}(\bm{k}|t)$ satisfy the symmetry relation $F_{nn'}(-\bm{k}|t)=F_{n'n}^*(\bm{k}|t)$. Moreover, the trace $Tr[\hat{D}_{l}^{pol}(-\bm{k})\hat{F}(\bm{k}|t)]$ is a real quantity, as a trace of a product of two self-adjoint operators. In summary, the first equation of motion (Eq.~\ref{eq:dsigma_ks/dt}) represent the equation for a harmonic oscillator driven by an external force acting from the side of the exciton quantum field.

It is noteworthy that the solution of Eqs.~(\ref{eq.:alpha_s(k)}) and (\ref{eq:dsigma_ks/dt}) for the electron-vibrational coupling parameter $\tilde{\alpha}_{l}(\bm{k};nn')$ without accounting for the initial condition is given by the integral
\begin{multline}
\tilde{\alpha}_{l}(\bm{k};nn')=2iD_{l}^{\ast pol}(\bm{k};nn')\\\times \int_{0}^{\infty}Tr[\hat{D}_{l}^{pol}(-\bm{k})\hat{F}(\bm{k}|t-\tau)]\exp(-\gamma\tau)\sin\left(  \omega_{l}\tau\right)  d\tau
\label{eq.:alpha^tilda_l(k)2}
\end{multline}

The second equation of motion was derived using a more general set-up. In other words, we take into account interaction of excitons with the light field. The interaction with light was modeled by introducing a non-Hermitian term into the total Hamiltonian. To evaluate the effects of the relaxation processes associated with the exciton-phonon interaction and the spontaneous emission rate, we used the non-crossing approximation~\cite{Fai07PRB}, which assumes that the exciton-vibration and exciton-photon interaction processes do not affect each other. This approximation is correct in the Condon approximation, when one assumes that the electronic transition is most likely to occur without changes in the positions of the nuclei. Within this approximation the non-Hermitian part can be described by introducing a decay term $-\hat{\Gamma}(C)$, which we specify later, into the resulting equation. Formally, to obtain the second equation of motion, one has to substitute the time-derivative $\dot{\sigma}_{\bm{k}}$ (Eq.~\ref{eq:dsigma_ks/dt}) into Eq.~(\ref{Cdot1}), which yields
\begin{multline}\label{eq:dC/dt}
\dot{C}_{n}(\bm{\sigma},\bm{q}) =-\hat{\Gamma}(C)+\sum_{n'\bm{k}}\tilde{\alpha}_{l}(\bm{k};n'n)C_{n'}(\bm{\sigma},\bm{q}+\bm{k})\\
 -i\left\{W_{n}(\bm{q})-\frac{1}{2}\sum_{\bm{k}}Tr[i\tilde{\alpha}_{l}(\bm{k})\hat{F}^{\dag}(\bm{k}|t)]\right\}C_{n}(\bm{\sigma},\bm{q}),
\end{multline}
where the term $\hat{\Gamma}(C)$ is the extra term introduced to describe the effective spontaneous decay of the given exciton state owing to interaction with the electromagnetic field. Note here, that the term ``$(1/2)\sum_{\bm{k}}Tr[i\tilde{\alpha}_{l}(\bm{k})F^{\dag}(\bm{k}|t)]$'', which stands in the curly braces, provides a nonlinear mean-field correction to the exciton eigenenergy $W_{n}(\bm{q})$, as this term is real (trace of two self-adjoint operators product) and independent of the wave vector $\bm{q}$. This equation is the second equation of motion for the model with an effective decay of the exciton.

Using Eq.(\ref{eq:dC/dt}), one can re-derive the equation for the evolution of the amplitude squared $|C_{n}(\bm{\sigma},\bm{q})|^{2}$ (population). Bearing in mind that $Tr[i\tilde{\alpha}_{l}(\bm{k})\hat{F}^{\dag}(\bm{k}|t)]$ is real we get
\begin{multline}
\frac{d}{dt}|C_{n}(\bm{\sigma},\bm{q})|^2
=-2\mathrm{Re}[C_{n}^*(\bm{\sigma},\bm{q}|t)\hat{\Gamma}(C)]\\
+2\mathrm{Re}\left[\sum_{n'\bm{k}}\tilde{\alpha}_{l}(\bm{k};n'n)C_{n}^*(\bm{\sigma},\bm{q})C_{n'}(\bm{\sigma},\bm{q}+\bm{k})\right]. \label{eq.:|C(q|t)|^2a}
\end{multline}
The second term on the right-hand side of Eq.~(\ref{eq.:|C(q|t)|^2a}) reproduces the right-hand side of Eq.~(\ref{eq:d|C(q'|t)|^2/dt}) and describes the relaxation in the $\bm{q}$-momentum space due to the exciton-phonon interaction, whereas the additional (first) term corresponds to the exciton probability decay.

Finally, we specify the spontaneous exciton decay term $\hat{\Gamma}(C)$ by entering Eqs.~(\ref{eq:dC/dt}), and (\ref{eq.:|C(q|t)|^2a}) for the quasi-2D  model. Consider a quasi-2D system (thin film) of thickness $l$ with the exciton wave function given by Eq.~(\ref{eq:2Dwavefunction}). In this case, the decay rate of the quasi-2D Wannier exciton with the lowest energy ($n=0$) and largest oscillator strength, and with the two-dimensional wavevector $\bm{q}$ collinear to the film plane, is given by~\cite{Hanamura88} (see also \cite{Liu_Lee80})
\begin{multline} \label{eq.:Gamma_q}
\Gamma_{\bm{q}}(\omega)=
\Gamma_{0}(\omega)\left(  \frac{\omega_{\bm{q}}}{\omega_{0}}\right)
^{2}\\
\times \left[  \sqrt{1-\frac{c^{2}q^{2}}{\omega^{2}}}\abs{\bm{e}_{\bm{k}}\times\hat{z}}^2 +\frac{c^{2}q^{2}}{\omega^{2}}\frac{1}{\sqrt{1-\frac{c^{2}q^{2}}{\omega^{2}}}}\abs{\bm{e}_{\bm{k}}\times\frac{\bm{q}}{q}}^2\right]
\end{multline}
for $\omega>cq$, and equals zero for $\omega<cq$. The factor $\Gamma_{0}(\omega)$ is
\begin{equation}
\Gamma_{0}(\omega)\equiv\Gamma_{\bm{q=}0}(\omega)=24\pi\left(
\frac{\lambda}{a_{0}}\right)  ^{2}\left(  \frac{\omega_{0}}{\omega}\right)
^{2}\gamma_{s} \label{eq.:Gamma_0}
\end{equation}
with $\gamma_{s}=\frac{4\abs{\bm{D}_{cv}}^{2}}{3\hbar\lambda^{3}}$, where $\lambda$ is the wavelength corresponding to the optical transition from the exciton state with the lowest energy ($n=0$), and $D_{cv}$ is the interband transition dipole moment. As it follows from Eq.(\ref{eq.:Gamma_0}), the radiative decay rate is enhanced by a factor of $24\pi\left(\frac{\lambda }{a_{0}}\right)^2$. This enhancement stems from the coherent nature of a 2D exciton with respect to the center-of-mass motion of the excitons. Finally, we can conclude that the decay term $\hat{\Gamma}(C)$ in Eqs.(\ref{eq:dC/dt}), and (\ref{eq.:|C(q|t)|^2a}) at $n'=0$ takes the simple form $\hat{\Gamma}(C)|_{n'=0}=\frac{1}{2}\Gamma_{\bm{q}}(\omega)C_{0}(\bm{\sigma},\bm{q}|t)$.

\section{Analysis of small deviations from the coherent superradiant state}
\label{sec:small_deviations}
Under perpendicular incidence of the pump field, excitons with $\bm{q}=0$ are created. It is known that for such zero-momentum excitons, the polariton effect vanishes, and their radiative decay rate is given by Eq.(\ref{eq.:Gamma_0})~\cite{Hanamura88}. As demonstrated in Section \ref{sec:molec_excitons}, such $\bm{q}=0$ state is a coherent state, allowing coherent superradiance, not just for Wannier excitons in polar crystals but also for molecular excitons.

Consider Eqs.(\ref{eq:dC/dt}) and (\ref{eq.:|C(q|t)|^2a}) for $n=0$. We will investigate the stability of the superradiant properties of the coherent state $\bm{q}=0$ (corresponding to $\bm{k}=0$ in the sums). Clearly, because the energy $\hbar W_{n}(\bm{k})$ scales as $k^{2}$, the superradiant state ($\bm{k}=0$) has the lowest energy. Moreover, the exciton-vibrational coupling parameter $\tilde{\alpha}_{l}(\bm{k};nn')$ equals zero at $\bm{k}=0$ (Eqs.(\ref{eq.:alpha^tilda_l(k)2}), (\ref{eq.:D^pol_l(k->0)2D}) and (\ref{eq.:D^pol_l(k;01)2D_2})). This behavior of the coupling parameter can be explained by the fact that for $k=0$ the macroscopic electric field of the LO phonon is uniform in space~\cite{Cardona2010}. Unlike uncorrelated electrons and holes that carry a net charge, the electrically neutral exciton does not interact with a uniform field. For a small non-zero $k$, the matrix elements $D_{l}^{pol}(\bm{k};00)$ and $D_{l}^{pol}(\bm{k};0(1,0))$ are proportional to $k$ (see Eq.~\ref{eq.:D^pol_l(k->0)2D}). This behavior is analogous to that of the matrix element for the optical quadrupole transitions. In other words, a state with a specific wave vector $\bm{q}$ behaves like an optically metastable state, when dipole transition is forbidden.

Consider Eq.(\ref{eq:dC/dt}) at $n=0$ and small $k$. Because of the absence of the ``vibration'' contribution to the coherent state attenuation, in the second order approximation over $\bm k$, Eqs.(\ref{eq.:alpha^tilda_l(k)2}) and (\ref{eq:dC/dt}) are reduced to
\begin{multline}
\tilde{\alpha}_{l}(\bm{k};nn')=2iD_{l}^{\ast pol}(\bm{k};nn')\\\times \int_{0}^{\infty}Tr[\hat{D}_{l}^{pol}(-\bm{k})\hat{F} (0|t-\tau)]\exp(-\gamma\tau)\sin\left(\omega_{l}\tau\right) d\tau
\label{eq.:alpha_l(small_k)a}
\end{multline}
\begin{multline}
\dot{C}_{0}(\bm{\sigma},\bm{q})=-iW_{0}(\bm{q})C_{0}
(\bm{\sigma},\bm{q}) \\
+\frac{1}{2}C_{0}(\bm{\sigma},\bm{q})\sum_{\bm{k},k<K}\tilde{\alpha}_{l}(\bm{k};00)-\frac{1}{2}
\Gamma_{\bm{q}}(\omega)C_{0}(\bm{\sigma},\bm{q})
\label{eq:dC/dt_polar2c}
\end{multline}
When deriving Eq.(\ref{eq:dC/dt_polar2c}), we took into account the smallness
of the non-diagonal terms $|D_{l}^{pol}(\bm{k};0(1,0))|^{2}$ with respect
to the diagonals $|D_{l}^{pol}(\bm{k};00)|^{2}$ (see Fig.\ref{fig:D2}) and the linear behaviour of $D_{l}^{pol}(\bm{k};00)\sim k$ at small $\bm k$ (Eq.~\ref{eq.:D^pol_l(k->0)2D}). Parameter $K$ is the limiting value for $k$ up to which the function $|D_{l}^{pol}(\bm{k};00)|^{2}$ can be approximated by $k^2$.
By comparing the graphs of the function  $|D_{l}^{pol}(\bm{k};00)|^{2}$ (Eq.~\ref{eq.:D^pol_l(k)2D}) with its first and second order expansions (see Fig.\ref{fig:D2}b and Eq.~\ref{eq.:D^pol_l(k->0)2D}) we found that the value of $K$ can be set to $1/a_{0}$. In other words, variations in state coherence over a scale shorter than $\frac{2\pi}{K}=2\pi a_{0}$ cannot be specified. Note that the procedure for limiting the possible wavevector values, $k<K$, is similar to the derivation of a block Hamiltonian using the Kadanoff transformation in the theory of phase transitions~\cite{Ma}.

The approximation used in this section leads to substitution of $F_{nn'}(\bm{k}|t)=\sum_{\bm{q}}\rho_{n,\bm{q}+\bm{k};n'\bm{q}}(t)$ by its value at $\bm{k}=0$, i.e. $F_{nn'}(0|t)=\sum_{\bm{q}}\rho_{n,\bm{q};n'\bm{q}}(t)$ (see Eqs.(\ref{eq.:alpha_l(small_k)a}) and (\ref{eq:dC/dt_polar2c})). In other words, we neglect the contribution of exciton-phonon interactions to the correlation of exciton states with different wave numbers. This approximation is valid provided that the neighboring electron-hole pairs are spatially separated by a distance, following the above estimation, at least $2\pi a_{0}$. Note that there an analogy exists between the approximation of small deviations from the coherent state and the van Hove theory of phase transitions, where the mode-mode couplings are also neglected~\cite{Ma}.

Solving Eq.(\ref{eq:dC/dt_polar2c}) where
\begin{equation}
\tilde{\alpha}_{l}(\mathbf{k};00)=2i|D_{l}^{pol}(\mathbf{k};00)|^{2}%
\frac{\omega_{l}}{\omega_{l}^{2}+\gamma^{2}}\label{eq.:alpha^tilda_l(k)3}%
\end{equation}
we get
\begin{multline}
C_{0}(\mathbf{\sigma},\mathbf{q}|t)=C_{0}(\mathbf{\sigma},\mathbf{q}%
|0)\exp\bigg\{-i\big[W_{0}(\mathbf{q})
\\-\sum_{\mathbf{k},k<K}\frac{\omega_{l} |D_{l}
^{pol}(\mathbf{k};00)|^{2}}{\omega_{l}^{2}+\gamma^{2}}\big]t-\frac{1}{2}%
\Gamma_{\mathbf{q}}(\omega)t\bigg\} \label{eq:C_0(q,t)}%
\end{multline}
One can show that $|C_{0}(\bm{\sigma},\bm{q})|^{2}=|C_{0}(\bm{\sigma},\bm{q}|0)|^{2}\exp[-\Gamma_{\bm{q}}(\omega)t]$. The last equation demonstrates that even for non-zero $k$, the "vibrational" contribution to the superradiant state attenuation is absent up to the order $k^2$ terms. This is true for a scale larger than $2\pi/K=2\pi a_{0}$.

\section{Analysis of the Semiclassical Hamiltonian}
\label{sec:SC_Hamiltonian}
The consideration of the coherent superradiant exciton state in Section \ref{sec:small_deviations} is limited to single-exciton approximation. In this regard, the following natural questions arise: `` What happens in the case of a multiple excited state? Why did we consider the superradiant state as the initial condition?'' Regarding the second question, one option for creating a coherent
superradiant state with $\bm{q}=0$ is the perpendicular incidence of the pump field (see the beginning of Section \ref{sec:small_deviations}). However, this issue requires more detailed consideration.

To answer these two questions, we consider the semiclassical version of the Hamiltonian $\hat{H}_{0}$, Eq.(\ref{eq.:H_03}). The semiclassical Hamiltonian is obtained by substituting the phonon operators $\hat{b}_{\bm{q}}^{\dag}$ and $\hat{b}_{\bm{q}}$ with the corresponding vibrational coherent states
$\sigma_{\bm{q}}^*$ and $\sigma_{\bm{q}}$, respectively:
\begin{multline}
\hat{H}_{0}^{SC}=\hbar\sum_{n\bm{q}}W_{n}(\bm{q})\hat{B}_{n\bm{q}}
^{\dag}\hat{B}_{n\bm{q}} \\
-i\hbar\sum_{nn'\bm{kq}}\tilde{\alpha}_{l}^*(\bm{k};nn')\hat{B}_{n,\bm{q}+\bm{k}}^{\dag
}\hat{B}_{n'\bm{q}}+\hbar\omega_{l}\sum_{\bm{\bm{q}}}
|\sigma_{\bm{q}}|^{2} \label{eq.:H^SC_0c}
\end{multline}
Considering only the lowest exciton state $n=0$ and considering the smallness of the non-diagonal terms $|D_{l}^{pol}(\bm{k};0\neq n)|^{2}$ with respect to the diagonals $|D_{l}^{pol}(\bm{k};00)|^{2}$, one can omit indices $n$ and $n'$ when writing operators $\hat{B}_{n\bm{q}}^{\dag}$ and $\hat{B}_{n'\bm{q}}$ in Eq.(\ref{eq.:H^SC_0c}). Note that in the theory of phase transitions, the wave number expansion is a usual approximation~\cite{Ma}. Bearing in mind Eq.(\ref{eq.:alpha^tilda_l(k)3}) and expansion (\ref{eq.:D^pol_l(k->0)2D}) up to terms of order $k^2$, the exciton-phonon interaction operator (the second term on the right-hand side of Eq.(\ref{eq.:H^SC_0c})) can be  approximately written as:
\begin{multline}
-i\hbar\sum_{\mathbf{kq}}\tilde{\alpha}_{l}^{\ast}(\mathbf{k};00)\hat{B}_{\mathbf{q}%
+\mathbf{k}}^{\dag}\hat{B}_{\mathbf{q}}\\
\approx -\frac{2\omega_{l}}{\omega_{l}^{2}%
+\gamma^{2}}\hbar\tilde{B}\sum_{\mathbf{q}}\hat{B}_{\mathbf{q}}^{\dag}\hat{B}_{\mathbf{q}%
}\sum_{\mathbf{k},k<K}k^{2}\label{eq.:H^pol(W)_I_SCsmall_k}%
\end{multline}
where $\tilde{B}=\frac{9\pi\omega^2_{l}\alpha}{256uV}(p_{h}^{2}-p_{e}^{2})^2 a_{0}^2$.  Comparison of Eq.(\ref{eq.:H^SC_0c}) with Eq.~(\ref{eq.:H^pol(W)_I_SCsmall_k}) leads to the conclusion that the presence of the vibrational coupling is reduced to the correction of the frequency $W_{0}(\bm{q})$ and does not violate the coherence of the initial state, up to the terms of order $k^2$. This conclusion generalizes the results of Section \ref{sec:small_deviations} beyond the framework of a single-exciton approximation. In particular, the series expansion up to the quadratic terms $k^2$ near the superradiant state ($\bm{q}=0$) corresponds to the absence of phase transition in the presence of exciton-phonon coupling, that is the maintenance of the superradiant state. It must be noted that in the case of single exciton the expansion to quadratic terms $k^2$ in the Hamiltonian leads to the neglect of the exciton-phonon interaction contribution into the correlation of exciton states with different wave numbers. The same expansion also allows us to neglect the exciton-exciton correlations induced by the exciton-phonon interaction in the case of multiple excitons, provided that neighboring excitons are separated by a distance of at least $2\pi a_{0}$.

To address the question of why the superradiant state is taken as an initial condition, we consider the exactly solvable model for long-wave excitons - the
Lipkin-Meshkov-Glick (LMG) model \cite{LMG1_1965,LMG2_1965,LMG3_1965}. We adopt the following convention for indices: Latin indices with subscript $e$
($h$), that is $m_{e}$ ($m_{h}$), represent electrons (holes). A particle (electron--hole pair) is denoted by a Latin index with no subscript $m=(m_{e}m_{h})$. Let us introduce the particle (electron-hole) operators
\begin{equation}
\hat{B}_{m_{e}m_{h}}^{\dag}=\hat{a}_{m_{e}}^{\dag}\hat{b}_{m_{h}}^{\dag}=\hat{B}_{m}^{\dag}\text{, \ }\hat{B}_{m_{e}m_{h}}=\hat{b}_{m_{h}}\hat{a}_{m_{e}}=\hat{B}_{m}
\end{equation}
where $\hat{a}_{m_{e}}$($\hat{a}_{m_{e}}^{\dag}$) and $\hat{b}_{m_{h}}$ ($\hat{b}_{m_{h}}^{\dag}$) are the operators for the electrons and holes, respectively. The LMG model assumes identical interactions between the different particles. Its Hamiltonian can be written as
\begin{equation}
\hat{H}_{LMG}=\hbar\sum_{m}\left(\hat{B}_{m}^{\dag}\hat{B}_{m}\bar{W}-|J|\sum_{m'\neq m}\hat{B}_{m}^{\dag}\hat{B}_{m'}\right)
\end{equation}
Define the energetic spin vector in the second quantization picture
\cite{Fai10PRB}%
\begin{equation}
\left(
\begin{array}
[c]{c}%
r_{1}\\
r_{2}\\
r_{3}%
\end{array}
\right)  =\left(
\begin{array}
[c]{c}%
B^{\dag}+B\\
i(B-B^{\dag})\\
\hat{n}_{2}-\hat{n}_{1}%
\end{array}
\right),  \label{eq.:Bloch}%
\end{equation}
where $\hat{n}_{2}=B^{\dag}B$ and $\hat{n}_{1}=BB^{\dag}$. Then $\hat{H}_{LMG}$ can be written as the Hamiltonian of the isotropic LMG model~\cite{Carrasco2016}:
\begin{equation}
\hat{H}_{LMG}^{iso}=-\hbar|J|\left(R_{1}^{2}+R_{2}^{2}-\frac{N}{2}I\right)-\hbar\bar{W}R_{3} \label{eq.:H^W_ex_sites3}%
\end{equation}
where $R_{i}=\sum_{m}r_{im}$ ($i=1,2,3$) are the components of the total energetic spin, $I$ is the unit operator, and the spectrum of $\hat{H}
_{LMG}^{iso}$ is independent of the sign before $\hbar\bar{W}R_{3}$. The eigenstates of the Hamiltonian $\hat{H}_{LMG}^{iso}$ are eigenstates
$\ket{R,M}$ of the operators $R^2$ and $R_{3}$.

The LMG model undergoes a second-order quantum phase transition at the value of parameter $h=\frac{\bar{W}}{N\abs{J}}=1$ \cite{Vidal2007}. Indeed, for $h\geq 1$ the minimum energy $E(M,R)$ is achieved when $R=M=N/2$, such that the ground state is the product of all spin-up states, which is similar to the ferromagnetic ground state. This corresponds to a completely inverted system, when the power of the spontaneous emission is proportional to $N$ \cite{Andreev93}. On the other hand, when $0\leq h<1$ the energy reaches a minimum when $S=N/2$ and $M$ is close to $\frac{\bar{W}}{2\abs{J}}\ll N$, so the ground state of $\hat{H}_{LMG}^{iso}$ is the totally symmetric Dicke state with $R=N/2$ and $M\ll N$ \cite{Carrasco2016}. We consider $N$ sufficiently large. Therefore, we implement the last case.

Considering the Fourier representation of the operators $\hat{B}_{m}=\frac{1}{\sqrt{N}}\sum_{\bm{q}}\hat{B}_{\bm{q}}\exp(i\bm{qR}_{m})$, $\hat{B}_{\bm{q}}=\frac{1}{\sqrt{N}}\sum_{m}\hat{B}_{m}\exp(-i\bm{q\cdot R}_{m})$ and Eq.(\ref{eq.:H^W_ex_sites3}), the isotropic LMG Hamiltonian $\hat{H}_{LMG}^{iso}$ can be rewritten in the momentum space. Substitution of the free exciton Hamiltonian $\hbar\sum_{\bm{q}}W(\bm{q})\hat{B}_{\bm{q}}^{\dag}\hat{B}_{\bm{q}}$ for $n=0$ in the semiclassical Hamiltonian $\hat{H}_{0}^{SC}$ (Eq.~\ref{eq.:H^SC_0c}) by $\hat{H}_{LMG}^{iso}$ and accounting for Eq.(\ref{eq.:H^pol(W)_I_SCsmall_k}), where we make the replacement $\sum_{\bm{q}}\hat{B}_{\bm{q}}^{\dag}\hat{B}_{\bm{q}}=\frac{N}{2}I+R_{3}$, gives:
\begin{multline}
\hat{H}_{0}^{SC}\approx-\hbar\abs{J}\left(R_{1}^{2}+R_{2}^{2}-\frac{N}{2}I\right)+\hbar\omega_{l}
\sum_{\bm{\bm{q}}}\abs{\sigma_{\bm{q}}}^{2}\\
-\hbar\left[\bar{W}+\frac{2\omega_{l}}{\omega_{l}^{2}+\gamma^{2}}\tilde
{B}\sum_{\mathbf{k},k<K}k^{2}\right]\left(\frac{N}{2}I+R_{3}\right) \label{eq.:H^SCsmall_k}
\end{multline}
In other words, the Wannier exciton-LO phonon interaction does not violate the isotropic LMG model up to terms of order $k^2$. In addition, the addition to $\bar{W}$ due to exciton-phonon interactions does not violate the previous criterion $h<1$ (see above). Therefore, the ground state of the Hamiltonian, Eq.(\ref{eq.:H^SCsmall_k}), remains the totally symmetric Dicke state.

\section{Consideration in terms of the density matrix}
\label{sec:DM}
Here we reformulate our theory by expressing it in terms of the density matrix, Eq.(\ref{eq:DM}). We consider the equations for the lowest exciton state ($n=n'=0$), i.e. we assume that exciton states other than $n=0$  are not excited. Then the expression for the density matrix is reduced to $\rho_{0\bm{q;}0\bm{q}'}(t) = C_{0}^{\ast}(\bm{\sigma}, \bm{q}') C_{0} (\bm{\sigma}, \bm{q})/\sum_{n\bm{q}''} \abs{C_{n}(\bm{\sigma}, \bm{q}'')}^2$. When we neglect the non-Hermitian part of the Hamiltonian, the trace of the density matrix is conserved owing to Eq.~(\ref{eq:conservation1}). Otherwise, the trace of the density matrix was not conserved. To resolve this issue, we supplemented the density matrix equations with an additional equation for the ground state (no excitons) population: $\dot{\rho}_{G\bm{q},G\bm{q}}(t)=\Gamma_{\bm{q}}(\omega)\rho_{0\bm{q},0\bm{q}}(t)$. The normalized sum $\sum_{n\bm{q}''}|C_{n}(\bm{\sigma},\bm{q}'')|^{2}$ can be replaced by a unit, assuming that all excited states with $n>0$ quickly decay into the $n=0$ state. Then $\rho_{0\bm{q},0\bm{q}'}$ reduces to $\rho_{0\bm{q},0\bm{q}'}=C_{0}^*(\bm{\sigma},\bm{q}') C_{0}(\bm{\sigma},\bm{q})$.

Eq. (\ref{eq.:|C(q|t)|^2a}) for $n'=0$ clearly represents the diagonal element $\rho_{0\bm{q},0\bm{q}}(t)$ of the density matrix
\begin{equation}
\dot{\rho}_{0\bm{q},0\bm{q}}(t)=2\operatorname{Re}\sum_{n\bm{k}}\tilde{\alpha}_{l}(\bm{k};n0)\rho_{n,\bm{q}+\bm{k};0\bm{q}}(t)-\Gamma_{\bm{q}}(\omega)\rho_{0\bm{q},0\bm{q}}(t)
\label{eq:drho_qq/dt}
\end{equation}
Eq.(\ref{eq:drho_qq/dt}) can be represented as a population master equation. First, using the Liouville operators~\cite{Zwa64}, Eq.~(\ref{eq:drho_qq/dt}) can
be rewritten as follows:
\begin{multline}
\dot{\rho}_{0\bm{q},0\bm{q}}(t)=-\int_{0}^{t}d\tau\sum_{n\bm{q}'\neq0\bm{q}}K_{0\bm{q},0\bm{q};n\bm{q}',n\bm{q}'}(\tau)\\\times
\Big[\rho_{n\bm{q}',n\bm{q}'}(t-\tau)-\rho_{0\bm{q},0\bm{q}}(t-\tau)\Big]-\Gamma_{\bm{q}}%
(\omega)\rho_{0\bm{q},0\bm{q}}(t) \label{eq:drho_qq/dtZwanzig}
\end{multline}
The kernel $K_{0\bm{q},0\bm{q};n\bm{q}',n\bm{q}'}(\tau)$ up to the second order in the exciton-phonon interaction parameter
$\tilde{\alpha}_{l}(\bm{k};n0)$ has the form
\begin{multline}
K_{0\bm{q},0\bm{q};n(\bm{q}+\bm{k)},n(\bm{q}+\bm{k)}
}(\tau)=2\mathrm{Re}\Big[e^{-i\omega_{n,\bm{q}+\bm{k};0\bm{q}}\tau}
\\\times \langle\tilde{\alpha}_{l}(\bm{k};n0|t)\tilde{\alpha}_{l}
(-\bm{k};0n|t-\tau)\rangle\Big], \label{eq:K_qq;q',q'}
\end{multline}
where $\omega_{n,\bm{q}+\bm{k};0\bm{q}}=W_{n}(\bm{q}+\bm{k})-W_{0}(\bm{q})$ and $\langle...\rangle$ denotes the thermal averaging. When the state of the phonon system does not depend on excitons, the quantum correlation function $\langle\tilde{\alpha}_{l}(\bm{k};n0|t)\tilde{\alpha
}_{l}(-\bm{k};0n|t-\tau)\rangle$ is given by 
\begin{multline}
\langle\tilde{\alpha}_{l}(\bm{k};n0|t)\tilde{\alpha}_{l}(-\bm{k}
;0n|t-\tau)\rangle\\=-|D_{l}^{pol}(\bm{k};n0)|^{2} \Big[\bar{n}_{l}e^{i\omega_{l}\tau}+(\bar{n}_{l}+1)e^{-i\omega_{l}\tau}\Big],
\label{eq:CorrFun(alpha_l)}
\end{multline}
where $\bar{n}_{l}=1/[\exp(\hbar\omega_{l}/k_{B}T)-1]$ is the thermally
averaged occupation number of the $l$-th mode.

To derive the equation for the density matrix one can also apply the non-crossing approximation \cite{Fai07PRB} and express
$\rho_{n,\bm{q}+\bm{k};0\bm{q}}(t)$ in terms of the populations (diagonal elements of the density matrix); thus, the equation for $\rho_{n,\bm{q}
+\bm{k};0\bm{q}}(t)$ can be obtained directly from Eq.(\ref{eq:dC/dt}).

In the Markovian approximation ($\tau$ is short) we can set the upper limit of the integration on the right-hand side of Eq.(\ref{eq:drho_qq/dtZwanzig}) equal to infinity and disregard the dependence of populations on a short time $\tau$. Then using Eqs.(\ref{eq:K_qq;q',q'}) and (\ref{eq:CorrFun(alpha_l)}), we obtain
\begin{multline}
\dot{\rho}_{0\bm{q},0\bm{q}}(t)   =2\pi\sum_{n\bm{k}}
\abs{D_{l}^{pol}(\bm{k};n0)}^{2}\\\times\Big[\bar{n}_{l}\delta(\omega_{n,\bm{q}
+\bm{k};0\bm{q}}-\omega_{l})+(\bar{n}_{l}+1)\delta(\omega
_{n,\bm{q}+\bm{k};0\bm{q}}+\omega_{l})\Big]\\
 \times\Big[\rho_{n,\bm{q}+\bm{k};n,\bm{q}+\bm{k}}
(t)-\rho_{0\bm{q},0\bm{q}}(t)\Big]-\Gamma_{\bm{q}}(\omega
)\rho_{0\bm{q},0\bm{q}}(t) \label{eq:drho_qq/dtMarkovian}
\end{multline}
Eq.(\ref{eq:drho_qq/dtMarkovian}) yields the same estimate of the LO phonons contribution to the lifetime of the $\bm{q}=0$ state as Eq.(3.1) in Ref.\cite{Toyozawa59}. From the arguments of the $\delta$-functions for the diagonal interaction ($n=0$) on the right-hand side of Eq.(\ref{eq:drho_qq/dtMarkovian}) we obtain expression for the wave number, $k=\frac{\sqrt{2M^*\omega_{l}}}{\sqrt{\hbar}}$, for the SF state ($\bm{q}=0$). To estimate the value of $k$ in terms of the exciton Bohr radius $a_{0}$, where $a_{0}=\hbar\varepsilon_{0}/(\alpha c\mu_{e}\mu_{h}M^*)$ and $\alpha=1/137$ is the fine-structure constant, we use the following values: $\varepsilon_{0}=10$, $\omega_{l}=100$ $cm^{\bm{-}1}$, $p_{h}=0.4$, $p_{e}=0.6$ and $M^*=0.4m_{e}=3. 64\cdot10^{- 28}g$. Substitution of these values gives $k=\frac{\sqrt{2M^*\omega_{l}}}{\sqrt{\hbar}}\approx2/a_{0}$. As shown in Fig.\ref{fig:D2_approximate}, in this region the function $|D_{l}^{pol}(\bm{k};00)|^{2}$ can be approximated as the sum of the quadratic and quaternary terms. Note that this approximation is analogous to the decay $\tilde{\Gamma}_{\bm{k}}$ for systems in which certain conservation laws require the constant part of $\tilde{\Gamma}_{\bm{k}}$ to be zero. For example, in Ref.\cite{Ma} the decay rate $\tilde{\Gamma}_{\bm{k}}$ was calculated using the Langevin equation. For an isotropic Heisenberg ferromagnet, the total spin ($\bm{k}=0$ mode) is conserved owing to spin rotational invariance. Thus, any noise cannot change the magnetization for $\bm{k}=0$ at any temperature, which requires the expansion $\tilde{\Gamma}_{\bm{k}}=ak^{2}+bk^{4}+O(k^{6})$ under the assumption that the total spin is conserved~\cite{Ma}. However, the equation for the diagonal density matrix is also an equation for the averaged population
operator. Van Kampen \cite{vanKampen1997} showed that the calculation of the averaged operator damping using the density matrix equation yielded the same
results as those obtained using the Langevin equation. Therefore, the arguments in Ref.\cite{Ma} supporting the stability of the $\bm{q}=0$ state are also applicable
to the superradiant state involving Frohlich interactions between the LO phonons and Wannier excitons.

The Markov equation for $\rho_{0\bm{q},0\bm{q}}$, Eq.(\ref{eq:drho_qq/dtMarkovian}), is related to the weak interaction described
in the perturbation theory framework. By contrast, the solution for $\tilde{\alpha}_{l}(\bm{k};nn')$, which includes the Hartree term,
Eq.(\ref{eq.:alpha^tilda_l(k)2}), corresponds to a strong interaction. In such a case, the mode-mode coupling terms can generate not only dissipative
motion but also orderly or systematic motion \cite{Ma}. This issue will be covered elsewhere.

\section{Perturbative solutions}
\label{sec:PT}

\subsection{Comparison with the model of $N$ identical two-level molecules
when the intermolecular Coulomb interactions are absent}
\label{sec:molec_excitons}
A model of identical two-level molecules describes the case of Frenkel excitons. Then the free exciton Hamiltonian in Eq.~(\ref{eq.:H_03}) is replaced as follows: $\hat{H}_{ex}^{W}\rightarrow\hat{H}_{ex}=\hbar\omega_{eg}\sum_{\bm{k}}\hat{B}_{\bm{k}}^{\dag}\hat{B}_{\bm{k}}$ with the eigenenergies $\hbar W(\bm{k})=\hbar\omega_{eg}=\hbar\omega_{el}+\sum_{\mu}\hbar\omega_{\mu}S_{\mu}$ corresponding to the frequency of the optical absorption spectrum maximum of
a molecule, and $\mu$ counts the optically active (OA) molecular vibrations with the frequency $\omega_{\mu}$ (vibrational modes that are reorganized after the electron excitation), $S_{\mu}=X_{\mu}^{2}$ is the standard Huang-Rhys factor, $X_{\mu}$ is a shift in the equilibrium position of the $\mu$-th OA vibration after the molecule electronic excitation~\cite{Fainberg22JPCA,Osipov_Fainberg23PRB}. The exciton-phonon interaction Hamiltonian takes the form%
\begin{equation}
\hat{H}_{I}=i\hbar\sum_{\mu\bm{kq}}D_{\mu}\hat{B}_{\bm{k+q}}^{\dag
}\hat{B}_{\bm{k}}(\hat{b}_{-\bm{q}\mu}^{\dag}+\hat{b}_{\bm{q}\mu}) \label{eq:H_I}%
\end{equation}
where $D_{\mu}=i\omega_{\mu}X_{\mu}N^{\bm{-}1/2}$ does not depend on $\bm{k}$. The hermiticity of $\hat{H}_{I}$, meaning that $\hat{H}_{I}=\hat
{H}_{I}^{\dag}$, requires the function $D_{\mu}$ to satisfy the symmetry relation: $D_{\mu}=-D_{\mu}^*$. In momentum representation
\begin{equation}
\ket{me}=\frac{1}{\sqrt{N}}\sum_{\bm{q}}\ket{\bm{q}} \exp(-i\bm{q\cdot r}_{m}), \label{eq.:|me>}%
\end{equation}
the superradiant state, which is  $\ket{SR}=\frac{1}{\sqrt{N}}\sum_{m=1}^{N}\ket{me}$ can be written as $|SR\rangle=\left\vert \bm{q=}0\right\rangle
$, and the effective non-Hermitian Hamiltonian is reduced to
\begin{equation}
\hat{H}_{eff}=\hbar\omega_{eg}\sum_{\bm{q}}\ket{\bm{q}}\bra{\bm{q}}-i\hbar\frac{N\Gamma}{2}\ket{\bm{q}}\bra{\bm{q}}\delta_{\bm{q0}}. \label{eq.:H_0(q)}
\end{equation}

The equations of motion for a single-exciton wavefunction for the model of two-level atoms can be similarly derived using the approach presented in Section~\ref{sec:Equations}. The resulting equation for the exciton wavefunction expansion coefficients is similar to Eq.(\ref{eq:dC/dt}), it takes the form
\begin{multline}
\dot{C}(\bm{\sigma},\bm{q})=-i\Big[\omega_{eg}-\frac{1}{2}
\sum_{\bm{k}}\alpha(\bm{k})F(\bm{k}|t)\Big]C(\bm{\sigma
},\bm{q})\\
-\frac{N\Gamma}{2}C(\bm{\sigma},\bm{q})\delta
_{\bm{q}0}-i\sum_{\bm{k}}\alpha(\bm{k})C(\bm{\sigma
},\bm{q}+\bm{k}) \label{eq:dC/dt3}
\end{multline}
Here $F(\bm{k}|t)=\sum_{\bm{q}}\rho_{\bm{q}+\bm{k},\bm{q}}(t)$, $\alpha(\bm{k})=\sum_{\mu}\alpha_{\mu}(\bm{k})$, and $\alpha(\bm{k})$ is associated with the electron-vibrational coupling parameter for molecule $m$, $\alpha_{m}=\sum_{\mu}\omega_{\mu}X_{\mu}(\sigma_{m,\mu}^*+\sigma_{m,\mu})$~\cite{Osipov_Fainberg23PRB}, through the Fourier expansion $\alpha(\bm{k})=\frac{1}{N}\sum_{m}\alpha_{m}e^{i\bm{kr}_{m}}$.

The quantity $\abs{C(\bm{\sigma},\bm{q}=0}^2$ represents the population evolution of the superradiant state, while $\abs{C(\bm{\sigma},\bm{q}\neq0|t)}^2$ describes the population evolution of the dark states. In this scenario, an equation similar to Eq.(\ref{eq.:|C(q|t)|^2a})
takes the following form%
\begin{multline}
\frac{d}{dt}|C(\bm{\sigma},\bm{q})|^{2}=2\operatorname{Im}
\sum_{\bm{k}}\alpha(\bm{k})C^*(\bm{\sigma},\bm{q})C(\bm{\sigma},\bm{q}+\bm{k})\\
 -N\Gamma\delta_{\bm{q}
0}\abs{C(\bm{\sigma},\bm{q})}^{2} \label{eq.:|C(q|t)|^2non}
\end{multline}
Substituting $\bm{q}=0$ into Eq.(\ref{eq:dC/dt3}) leads to the equation for the superradiant state. Conversely, the equations for $\bm{q\neq}0$ describe the dark state evolution.

Parameter $\alpha_{m}(t)$ can be considered as a random Gaussian process for low frequency OA vibrations \cite{Fainberg19JPCC,Fainberg22JPCA}. In a
particular case, this process can be a Gauss-Markov process with an exponential correlation function and a characteristic attenuation time $\tau_{c}$,
\begin{equation}
K_{m}(t'-t^{\prime\prime})\equiv\langle\bar{\alpha}_{m}(t')\bar{\alpha}_{m}(t^{\prime\prime})\rangle=K_{m}(0)e^{-|t'-t''|/\tau_{c}} \label{eq:K_m(t)}%
\end{equation}
where $\bar{\alpha}_{m}=\alpha_{m}(t)-\langle\alpha_{m}\rangle$ are the centered quantities. The mathematical formalism of this model resembles the disorder effects caused by the interaction between molecules and their surrounding environment. In a recent study conducted by Nitzan et al. \cite{Nitzan22PRA}, the phenomenon of superradiance in disordered molecular ensembles was explored  through both numerical and analytical methods, with a focus on perturbation analysis of the influence of disorder effects. The study exclusively examined single excitation states, while Coulomb intermolecular interactions were disregarded. The analytical findings in Ref.\cite{Nitzan22PRA} can be derived from Eqs.(\ref{eq:dC/dt3}) and (\ref{eq:K_m(t)}) for the averaged values $\langle\abs{C(\bm{\sigma},\bm{q}|t)}^2\rangle$, as detailed in Appendix \ref{AppB}. As demonstrated in the aforementioned reference, the presence of inhomogeneous broadening leads to a loss of coherence in the collective molecular excitation and consequent suppression of superradiant emission. This outcome arises from the fact that, in the model of $N$ identical two-level molecules, the quantities $|D_{\mu }(\bm{k})|^{2}=\omega_{\mu}^{2}S_{\mu}/N$ remain constant and independent of $\bm{k}$, in contrast to the scenario involving Frohlich interactions between LO phonons and Wannier excitons.

\subsection{Frohlich interactions of LO phonons with Wannier excitons in polar
crystals}
\label{sec:FrohlichPT}
In this section we present the perturbative solution to  Eqs.(\ref{eq:dC/dt}) and (\ref{eq.:|C(q|t)|^2a}) over the exciton-phonon interaction in the model of the Wannier excitons in polar crystals. The exciton expansion coefficient is represented in the following form:
\begin{equation}
C_{n}(\bm{\sigma},\bm{q})\approx C_{n}^{(0)}(\bm{\sigma
},\bm{q})+C_{n}^{(1)}(\bm{\sigma},\bm{q}),
\label{eq.:C(q|t)=C^(0)+C^(1)a}
\end{equation}
where
\begin{multline}
C_{n}^{(0)}(\bm{\sigma},\bm{q})=\delta_{n_{0}}\delta_{\bm{q} 0}C_{0}^{(0)}(\bm{\sigma},0|t=0)e^{-iW_{0}(\bm{q}=0)t-\frac{1}{2}\Gamma_{0}(\omega)t} \label{eq.:C^(0)(q,t)2a}
\end{multline}
(see also Eqs.(\ref{eq.:C(q|t)=C^(0)+C^(1)}) and (\ref{eq.:C^(0)(q,t)2}) of
Appendix~\ref{AppB}). Here we assume that only the superradiant state with
$\bm{q}=0$ is initially excited.

According to Eq.(\ref{eq:dC/dt}), $C^{(1)}(\bm{\sigma},\bm{q}|t)$ obeys the equation
\begin{multline}
\dot{C}_{0}^{(1)}(\bm{\sigma},\bm{q})+i\tilde{W}_{0}(\bm{q}
)C_{0}^{(1)}(\bm{\sigma},\bm{q})=i\sum_{\bm{k}}C_{0}^{(0)}(\bm{\sigma},\bm{q}=0)
\\\times \left\{\frac{1}
{2}Tr[i\tilde{\alpha}_{l}^{(1)}(\bm{k})\hat{F}^{(0)\dag}(\bm{k}
|t)]-\tilde{\alpha}_{l}^{(1)}(\bm{k};00)\delta_{\bm{q},-\bm{k}
}\right\}, \label{eq.:C^(1)(q,t)2}
\end{multline}
where $\tilde{W}_{0}(\bm{q})=W_{0}(\bm{q})-\frac{i}{2}\Gamma
_{\bm{q}}(\omega)$, $F_{nn'}^{(0)}(\bm{k}|t)=\delta
_{nn'}\delta_{n0}\delta_{\bm{k}0}$, $Tr[i\tilde{\alpha}_{l}%
^{(1)}(\bm{k})\hat{F}^{(0)\dag}(\bm{k}|t)]=i\tilde{\alpha}_{l}%
^{(1)}(\bm{k};00)\delta_{\bm{k}0}$, and
\begin{equation}
\tilde{\alpha}_{l}^{(1)}(\mathbf{k};00)=2i|D_{l}^{pol}(\mathbf{k}%
;00)|^{2}\delta_{\mathbf{k}0}\frac{\omega_{l}}{\omega_{l}^{2}+\gamma^{2}%
}\label{eq.:alpha^(1)_l(k)}%
\end{equation}
 Because $F_{nn'}^{(0)}(\bm{k},t)\sim\delta_{\bm{k}0}$, argument $\bm{k}$ of $\tilde{\alpha}_{l}^{(1)}(\bm{k};00)$ equals zero. Consequently, $\tilde{\alpha}_{l}^{(1)}(\bm{k}=0;00)=0$ because $|D_{l}^{pol}(\bm{k}=0;00)|^{2}=0$. This immediately nullifies the right-hand side of Eq.(\ref{eq.:C^(1)(q,t)2}). As a result, in the first approximation, $C_{0}^{(1)}(\bm{\sigma},\bm{q})=0$ for the initial condition $C_{0}^{(1)}(\bm{\sigma},\bm{q}|t=0)=0$. It can be demonstrated that higher-order coefficients are also zero (see Appendix~\ref{AppC}). This is in contrast to the model of $N$ identical two-level molecules (see Section \ref{sec:molec_excitons} and Appendix~\ref{AppB}), where the interaction coefficient is independent of $\bm{k}$ constant, $D_{\mu}=i\omega_{\mu}X_{\mu}N^{-1/2}$, and the electron-vibrational coupling, $\alpha_{\mu}(\bm{k})$, obeys the equation
\begin{equation}
\alpha_{\mu}(\mathbf{k})=2\frac{\omega_{\mu}^{2}S_{\mu}}{N}\int_{0}^{\infty
}d\tau F^{\ast}(\mathbf{k}|t-\tau)\exp(-\gamma\tau)\sin\left(  \omega_{\mu
}\tau\right)  \label{eq.:alpha_mu(k)}%
\end{equation}
In this case, $\alpha(\bm{k}=0)=\sum_{\mu }\alpha_{\mu}(\bm{k}=0)=\frac{1}{N}\sum_{m}\alpha_{m}\neq0$ is the average parameter of the electron-vibrational coupling. Therefore, $\alpha^{(1)}(\mathbf{k})=\sum_{\mu}\alpha_{\mu}^{(1)}%
(\mathbf{k})=2\frac{\omega_{\mu}^{2}S_{\mu}}{N}\int_{0}^{\infty}d\tau
F^{(0)\ast}(\mathbf{k}|t-\tau)\exp(-\gamma\tau)\sin\left(  \omega_{\mu}%
\tau\right)  \neq0$, and the first-order coefficients $C^{(1)}(\bm{\sigma},\bm{q})$ differ from zero (see Appendix~\ref{AppB}).
\newline
\par
In summary, perturbation theory calculations provide additional evidence supporting the stability of the superradiant properties of the $\bm{q}=0$ coherent state in hybrid perovskite thin films, with respect to the perturbation caused by the LO phonon-exciton Frohlich interaction. As discussed in Section \ref{sec:DM}, there are systems where certain conservation laws force the $\bm k$-independent part of the decay rate $\tilde{\Gamma}_{\bm{k}}$ to become zero~\cite{Ma}. Our consideration generalizes this criterion to the case of the zero-constant part of the matrix element $D_{l}^{pol}(\bm{k};n0)$ (Eq.~\ref{eq.:D^pol_l(k->0)2D}), when the symmetrical state, $\bm{k}=0$ is conserved. Note that the dumping rate $\tilde{\Gamma}_{\bm{k}}\sim
\sum_{n}|D_{l}^{pol}(\bm{k};n0)|^{2}$, which bridges the two considerations.

\section{Conclusion}\label{sec:Conclusion}
This study provides insights into the mechanism of high-temperature SF in hybrid perovskite thin films. SF was observed at 78 K in methyl ammonium lead iodide perovskite (MAPbI$_{3}$) thin films \cite{Gundogdu2021Nature_Phot} and at room temperature in quasi-two-dimensional phenethylammonium cesium lead bromide perovskite (PEA:CsPbBr$_{3}$) \cite{Gundogdu2022Nature_Phot}. In both materials, the interaction between the Wannier excitons and LO phonons arises from the Frohlich interaction. We considered a quasi-2D Wannier exciton in a thin film interacting with LO phonons in polar crystals, and calculated the interaction Hamiltonian characterized by a strong wave vector dependence. Using the multiconfiguration Hartree approach, we derived the equations of motion for the Wannier exciton wavefunction, with the vibration degrees of freedom interacting with the exciton via the mean-field Hartree term. The LO phonon-exciton Frohlich interaction vanishes for the superradiant state at the 
zero-wave vector. For small deviations, a new quadratic in the deviation state emerges that retains superradiant properties on a scale greater than $2\pi/K=2\pi a_{0}$. In other words, the superradiant properties of the coherent state in hybrid perovskites are stable against perturbations caused by the LO phonon-exciton Frohlich interaction. We generalized this finding beyond the framework of a single-exciton approximation by considering a multiple excited state using a semiclassical Hamiltonian. We also considered an exactly solvable model for long-wave excitons, the LMG model \cite{LMG1_1965,LMG2_1965,LMG3_1965}, the ground state of which is the totally symmetric Dicke state. In this case, the series expansion to quadratic terms $k^2$ near the ground state corresponds to the absence of a phase transition in the presence of exciton-phonon coupling, that is, the maintenance of the superradiant state. A key requirement for high-temperature SF in hybrid perovskite thin films is the formation of Wannier excitons coupled to the LO phonon via Frohlich interaction. This can be understood as follows: at zero wave vector $\bm{k}$, the macroscopic LO phonon electric field is uniform in space and thus cannot alter the energy of neutral excitons, unlike for the uncorrelated electrons and holes, which carry a net charge. The Wannier exciton-LO phonon coupling $D_{l}^{pol}(\bm{k};nn')$ contains no wave-vector independent term. For small non-zero wave vectors, the coupling $D_{l}^{pol}(\bm{k};nn')$ is proportional to $k$, analogous to the matrix element for optical quadrupole transitions. A state with a specific wave vector $\bm{q}$ can be metastable, similar to a metastable state that can occur in optics when no dipole transition is allowed.

We rewrote our theory in terms of the density matrix, which does not contain a highly nonlinear Hartree term in the 
wave-function equation. Therefore, the density matrix equations may be simpler than the corresponding wavefunction equations. This density matrix-based consideration allowed us to draw an analogy between our system and stable systems where certain conservation laws forced the constant part of the decay $\tilde{\Gamma}_{\bm{k}}$ to zero \cite{Ma}. We associate this criterion with the absence of a constant component in the matrix elements $D_{l}^{pol}(\bm{k};nn')$ (including $D_{l}^{pol}(\bm{k};00)$), when the symmetrical superradiant state $\bm{k}=0$ is conserved. Perturbation theory calculations provide a good illustration of this rule. Indeed, the electron-vibrational interactions in the model of $N$ identical two-level molecules, where the matrix elements $\abs{D_{\mu (\bm{k})}}^{2}=\omega_{\mu}^{2}S_{\mu}/N$ are independent of $\bm{k}$,  destroy the coherence of the collective molecular excitations and, consequently, suppress SF. In contrast, the LO phonon-exciton Frohlich interactions, for which $\abs{D_l^{pol}(\bm{k};nn')}^2$ (and $\abs{D_l^{pol}(\bm{k};00)}^2$) have no constant component, do not disrupt the superradiant coherent state.

To summarize, the present work elucidates the conditions for maintaining the superradiant properties of the coherent state at high temperatures, which can inform the design of new quantum technology systems. However, several unresolved problems remain. First, the
problem of the influence of LO phonon-exciton Frohlich interactions on the superradiant properties on a scale of the order or smaller than $2\pi/K=2\pi a_{0}$, that is beyond the series expansion to quadratic terms $k^2$ near the superradiant state. Such an extension will enable us to study exciton-exciton correlations and the correlation of exciton states with different wave numbers induced by exciton-phonon interactions. In addition, the Markovian master equation (\ref{eq:drho_qq/dtMarkovian}) describes the attenuation for weak LO phonon-exciton interactions. This weak interaction leads to the mode-mode coupling in terms of the theory of critical phenomena \cite{Ma}. For stronger interactions, the solution of $\tilde{\alpha}_{l}(\bm{k};nn')$ with the Hartree term in, Eq.(\ref{eq.:alpha^tilda_l(k)2}), has to be found. In cases of stronger interaction, the mode-mode coupling terms may lead to regular or organized motion. The problems associated with strong interactions will be covered separately.
\section{Acknowledgments}

We thank Kenan Gundogdu for the useful discussions.

%

\appendix
\section{Matrix elements of Hamiltonians}\label{AppA}

To continue our calculations, we utilize the general observation that the matrix
elements $\langle\bm{\sigma}',\bm{q}'|\hat{H}%
_{0}|\bm{\sigma},\bm{q}\rangle$ of a generic Hamiltonian $\hat{H}_{0}%
$, normally ordered in $\hat{b}_{\bm{q}s}^{\dag}$ and $\hat{b}_{\bm{q}s}$, can be
obtained by replacing the operators $\hat{b}_{\bm{q}s}^{\dag}$ and
$\hat{b}_{\bm{q}s}$ with ${\sigma_{\bm{q},s}'}^*$ and
$\sigma_{\bm{q},s}$, respectively. That is:
\begin{equation}
\langle\bm{\sigma}',n'\bm{q}'|\hat{H}%
_{0}|\bm{\sigma},n\bm{q}\rangle=\langle\bm{\sigma}^{\prime
}|\bm{\sigma}\rangle\langle n'\bm{q}'|\hat{H}%
_{0}(\bm{\sigma}^{\prime\ast},\bm{\sigma})|n\bm{q}\rangle
\label{HamPol}%
\end{equation}
And correspondingly:
\begin{multline}
\langle\bm{\sigma}',n'\bm{q}'|\hat{b}_{\bm{k}%
}\hat{H}_{0}|\bm{\sigma},n\bm{q}\rangle=\sigma_{\bm{k}}%
\langle\bm{\sigma}',n'\bm{q}'|\hat{H}%
_{0}|\bm{\sigma},n\bm{q}\rangle\\
+\langle\bm{\sigma}^{\prime
}|\bm{\sigma}\rangle\frac{\partial}{\partial{\sigma_{\bm{k}}%
^{\prime\ast}}}\langle n'\bm{q}'|\hat{H}_{0}%
(\bm{\sigma}^{\prime\ast},\bm{\sigma})|n\bm{q}\rangle\label{cmHam}%
\end{multline}
Therefore, we obtain:
\begin{equation}
\frac{i}{\hbar}\langle\bm{\sigma}',n'\bm{q}'|\hat{H}_{ex}^{W}|\bm{\sigma},n\bm{q}\rangle=i\langle\bm{\sigma
}'|\bm{\sigma}\rangle W_{n}(\bm{q}')\delta
_{nn'}\delta_{\bm{qq}'} \label{eq:<q'|H_ex|q>}
\end{equation}
and%
\begin{multline}
\frac{i}{\hbar}\langle\bm{\sigma}',n'\bm{q}^{\prime
}|\hat{H}_{ph}|\bm{\sigma},n\bm{q}\rangle=i\langle\bm{\sigma
}'|\bm{\sigma}\rangle\delta_{nn'}\delta_{\bm{qq}%
'}\omega_{l}\\\times \sum_{\bm{k}'}[({\sigma_{\bm{k}'%
}^{\prime\ast}-}\sigma_{\bm{k}'}^*)\sigma_{\bm{k}%
'}+\sigma_{\bm{k}'}^*\sigma_{\bm{k}'}]
\label{eq:<q'|H_ph|q>}
\end{multline}
Using Eq.~(\ref{cmHam}), we also derive
\begin{equation}
\frac{i}{\hbar}\langle\bm{\sigma}',n'\bm{q}^{\prime
}|\hat{b}_{\bm{k}}\hat{H}_{ex}^{W}|\bm{\sigma},n\bm{q}\rangle
=i\braket{\bm{\sigma}'|\bm{\sigma}}\sigma_{\bm{k}%
}W_{n}(\bm{q}')\delta_{nn'}\delta_{\bm{qq}'}
\label{eq:<q'|b_ksH_ex|q>}%
\end{equation}
and%
\begin{multline}
\frac{i}{\hbar}\langle\bm{\sigma}',n'\bm{q}^{\prime
}|\hat{b}_{\bm{k}}\hat{H}_{ph}|\bm{\sigma},n\bm{q}\rangle=i\langle
\bm{\sigma}'|\bm{\sigma}\rangle\delta_{nn'}%
\delta_{\bm{qq}'}\sigma_{\bm{k}}\omega_{l}\\\times
\left\{1+\sum
_{\bm{k}'}[({\sigma_{\bm{k}'}^{\prime\ast}-}%
\sigma_{\bm{k}'}^*)\sigma_{\bm{k}'}%
+\sigma_{\bm{k}'}^*\sigma_{\bm{k}'}]\right\}
\label{eq:b_ksH_ph}%
\end{multline}
To obtain the expressions for the interaction
term we, first, calculate the ``sandwich'' $\langle n'\bm{q}'|\hat{B}_{n'',\bm{k+k}'}^{\dag}\hat{B}_{n'''\bm{k}}|n\bm{q}\rangle=\delta_{n'n^{\prime\prime}}%
\delta_{n^{\prime\prime\prime}n}\delta_{\bm{q}',\bm{k+k}%
'}\delta_{\bm{kq}}$ . This gives
\begin{multline}
-\frac{i}{\hbar}\langle\bm{\sigma}',n'\bm{q}^{\prime
}|\hat{H}_{I}^{W}|\bm{\sigma},n\bm{q}\rangle=\langle\bm{\sigma
}'|\bm{\sigma}\rangle D_{l}^{pol}(\bm{q}'%
\bm{-q};n'n)\\\times \left[({\sigma'}_{\bm{q-q}'}^{\ast
}-{\sigma}_{\bm{q-q}'}^*)+({\sigma}_{\bm{q-q}'%
}^*-\sigma_{\bm{q}'\bm{-q}})\right] \label{eq:<q'|H_I|q>}%
\end{multline}
Consequently, we also have
\begin{multline}
-\frac{i}{\hbar}\langle\bm{\sigma}',n'\bm{q}^{\prime
}|\hat{b}_{\bm{k}}\hat{H}_{I}^{W}|\bm{\sigma},n\bm{q}\rangle
=\braket{\bm{\sigma}'|\bm{\sigma}}\\\times
\Big\{\sigma_{\bm{k}%
}D_{l}^{pol}(\bm{q}'\bm{-q};n'n)[({\sigma'%
}_{\bm{q-q}'}^*-{\sigma}_{\bm{q-q}'}^{\ast
})+({\sigma}_{\bm{q-q}'}^*-\sigma_{\bm{q}^{\prime
}\bm{-q}})] \\
  +D_{l}^{\ast pol}(\bm{k};nn')\delta_{\bm{k},\bm{q-q}%
'}\Big\}. \label{eq:<q'|b_ksH_I|q>}%
\end{multline}

\section{ Comparison with Ref. \cite{Nitzan22PRA}.}\label{AppB}
The analytical results in Ref.\cite{Nitzan22PRA} can be derived from Eqs.(\ref{eq:dC/dt3}) and (\ref{eq:K_m(t)}) for the average values
$\langle|C(\bm{\sigma},\bm{q})|^{2}\rangle$. To obtain these results, we solved Eq.(\ref{eq:dC/dt3}) using perturbation theory with
respect to $\alpha(\bm{k})$:
\begin{equation}
C(\bm{\sigma},\bm{q})\approx C^{(0)}(\bm{\sigma},\bm{q}
)+C^{(1)}(\bm{\sigma},\bm{q}) \label{eq.:C(q|t)=C^(0)+C^(1)}%
\end{equation}
where
\begin{equation}
C^{(0)}(\bm{\sigma},\bm{q})=C^{(0)}(\bm{\sigma},\bm{q}%
=0|t=0)e^{-i\omega_{eg}t-\frac{N\Gamma}{2}t} \label{eq.:C^(0)(q,t)2}%
\end{equation}
We assumed that initially, only the superradiant state was excited.

The function $C^{(1)}(\bm{\sigma},\bm{q})$ obeys the equation
\begin{multline}
\dot{C^{(1)}}(\bm{\sigma},\bm{q})=-i\omega_{eg}C^{(1)}%
(\bm{\sigma},\bm{q})-\frac{N\Gamma}{2}C^{(1)}(\bm{\sigma
},\bm{q})\delta_{\bm{q}0}\\
-i\sum_{\bm{k}}\alpha(\bm{k}%
)\left[C^{(0)}(\bm{\sigma},\bm{q}+\bm{k})-\frac{1}{2}C^{(0)}
(\bm{\sigma},\bm{q})F^{(0)}(\bm{k},t)\right],
\label{eq.:dC^(1)(sigma,q)/dt}%
\end{multline}
where $F^{(0)}(\bm{k},t)=\sum_{\bm{q}}\rho_{\bm{q}+\bm{k}%
,\bm{q}}^{(0)}(t)=\delta_{\bm{k}0}$.

The solution for Eq.(\ref{eq.:dC^(1)(sigma,q)/dt}) can be expressed as:
\begin{multline}
C^{(1)}(\bm{\sigma},\bm{q})   =iC^{(0)}(\bm{\sigma}
,0|t=0)e^{-(i\omega_{eg}+\frac{N\Gamma}{2})t}\\\times
\int_{0}^{t}dx
\Big[-\alpha(-\bm{q},t-x)e^{\frac{N\Gamma}{2}(1-\delta_{\bm{q}0})x} \\
   +\frac{1}{2}\delta_{\bm{q}0}e^{\frac{3N\Gamma}{2}x-\frac{N\Gamma}{2}\delta_{\bm{q}0}x-N\Gamma t}
   \alpha(\bm{q}=0,t-x)\Big]
\label{eq.:C^(1)(q,t)}%
\end{multline}
assuming $C^{(1)}(\bm{\sigma},\bm{q}|t=0)=0$.

The superradiant state ($\bm{q}=0$) coefficient $C^{(1)}(\bm{\sigma },\bm{q}=0)$ matches the results in Ref. \cite{Nitzan22PRA} if the
Hartree term (the second term on the right-hand side of Eq.(\ref{eq.:C^(1)(q,t)})) is neglected. Substituting the expansion $\alpha
(-\bm{q},t)=\frac{1}{N}\sum_{m}\alpha_{m}(t)\exp(\bm{-}i\bm{qr}_{m})$ into Eq.(\ref{eq.:C^(1)(q,t)}) gives:
\begin{multline}
C^{(1)}(\bm{\sigma},\bm{q} =0)=-\frac{i}{N}C^{(0)}
(\bm{\sigma},0|t=0) e^{-i\omega_{eg}t}\\
\int_{0}^{t}dt' \Big \{e^{-\frac{N\Gamma}{2}t}
\sum_{m}\alpha_{m}(t')
 -\frac{\delta_{\bm{q}0}}{2}e^{-\frac{N\Gamma}{2}t}
\sum_{m}\alpha_{m}(t')e^{-N\Gamma t'}\Big\}
\label{eq.:C^(1)(q=0,t)2}
\end{multline}
The first term on the right-hand side of Eq.(\ref{eq.:C^(1)(q=0,t)2}) is
equivalent to that in Eq.(B6) in Ref. \cite{Nitzan22PRA} if we set $C^{(0)}
(\bm{\sigma},0|t=0)$ equal to 1 and treat $\alpha_{m}$ as a centered variable.

For the dark states ($\bm{q}\neq0$), Eq.(\ref{eq.:C^(1)(q,t)}) provides
the following result:
\begin{multline}
C^{(1)}(\bm{\sigma},\bm{q}\neq 0)=-iC^{(0)}(\bm{\sigma}
,0|t=0)e^{-i\omega_{eg}t}\\\times
\int_{0}^{t} \frac{dt'}{N}\sum_{m}\alpha_{m}(t')e^{\bm{-}i\bm{qr}_{m}}%
e^{-\frac{N\Gamma}{2}t'} \label{eq.:C^(1)(q,t)Dark}%
\end{multline}
Moving to the probability densities $\left\vert C^{(1)}(\bm{\sigma
},\bm{q}\neq0)\right\vert ^{2}$, summing over dark states ($\bm{q}%
\neq0$), and averaging over $\alpha_{m}$, we obtain:%
\begin{multline}
\sum_{\bm{q}\neq0}
\langle\left\vert C^{(1)}(\bm{\sigma},\bm{q}\neq0)\right\vert
^{2}\rangle=\frac{N-1}{N^{2}}
\\\times \int_{0}^{t}\int_{0}^{t}
\sum_{m}\langle\alpha_{m}(t')\alpha_{m}(t^{\prime\prime})\rangle
e^{-\frac{N\Gamma}{2}(t'+t'')}dt'
dt'' \label{eq.:SumC^(1)_k}%
\end{multline}
where we used $\sum_{\bm{q\neq}0}
\exp[\bm{-}i\bm{q}(\bm{r}_{m}-\bm{r}_{j})]=(N-1)\delta_{jm}$,
with $N-1$ being the number of dark states.

Since $\langle\left\vert C^{(1)}(\bm{\sigma},\bm{q}\neq0)\right\vert
^{2}\rangle$ does not depend on the dark state index \cite{Nitzan22PRA}, then
$\langle\left\vert C^{(1)}(\bm{\sigma},\bm{q}\neq0)\right\vert
^{2}\rangle=\frac{1}{N - 1}\sum_{\bm{q}\neq0}
\langle\left\vert C^{(1)}(\bm{\sigma},\bm{q}\neq0)\right\vert
^{2}\rangle$. By assuming $\alpha_{m}$ to be a centered quantity, utilizing Eq.(\ref{eq:K_m(t)}), we obtain from Eq.(\ref{eq.:SumC^(1)_k}):
\begin{multline}
\langle\left\vert C^{(1)}(\bm{\sigma},\bm{q}\neq0)\right\vert ^{2}\rangle\\ =\frac{K(0)}{N}\int_{0}^{t}\int_{0}^{t}
e^{-\frac{N\Gamma}{2}(t'+t'')}e^{-\frac{|t'-t''|}{\tau_{c}}}dt'dt''
\label{eq.:|C^(1)_k|^2}%
\end{multline}
In the derivation of Eq.(\ref{eq.:|C^(1)_k|^2}), we made the assumption, 
stated in Ref.\cite{Nitzan22PRA}, that the correlation functions,
Eq.(\ref{eq:K_m(t)}), do not depend on $m$. As a result, we obtain:%
\begin{multline}
\sum_{m}\langle\alpha_{m}(t')\alpha_{m}(t'')\rangle
=\sum_{m}K_{m}(0)e^{-\frac{|t'-t''|}{\tau_{c}}}\\=NK(0)e^{-\frac{|t'-t''|}{\tau_{c}}},
\end{multline}
where $K_{m}(0)=K(0)$. Eq.(\ref{eq.:|C^(1)_k|^2}) corresponds precisely to
Eq.(B.12) in Ref.\cite{Nitzan22PRA}.

Thus, the analytical results in Ref.\cite{Nitzan22PRA} for coefficients $C(\bm{\sigma},\bm{q})$ can be derived using our theory by disregarding the Hartree term in the equation for the coefficients of a single-exciton wavefunction.\textbf{ }It is important to highlight that the terms
of this nature were also eliminated from Eq.(\ref{eq.:|C(q|t)|^2non}) for the population $|C(\bm{\sigma},\bm{q})|^{2}$. As the main objective
of Ref.\cite{Nitzan22PRA} was to calculate the populations, the absence of the Hartree term in the expressions for the amplitudes $C(\bm{\sigma
},\bm{q})$ was not significant.

\section{Calculation of coefficients using the Poisson method.}\label{AppC}
The calculation of $C^{(1)}(\bm{\sigma},\bm{q})$ follows a method similar
 to the first approximation calculation in the Poisson method
\cite{Bogoliubov61}. Eq.(\ref{eq:dC/dt}) can be expressed as follows:
\begin{equation}
\dot{C}_{0}(\bm{\sigma},\bm{q})+i\tilde{W}_{0}(\bm{q}%
)C_{0}(\bm{\sigma},\bm{q})=\varepsilon f\left(  C\right)  ,
\label{eq:P1}%
\end{equation}
where the function $f\left(  C\right)  $, which is proportional to the small
parameter $\varepsilon$, represents the right-hand side of Eq.(\ref{eq:dC/dt}%
). In the Poisson method, the solution is sought as a series
\begin{multline}
C_{0}(\bm{\sigma},\bm{q})=C_{0}^{(0)}(\bm{\sigma},\bm{q}%
)+\varepsilon C_{0}^{(1)}(\bm{\sigma},\bm{q})\\
+\varepsilon^{2}%
C_{0}^{(2)}(\bm{\sigma},\bm{q})+... \label{eq:P2}%
\end{multline}
By substituting series (\ref{eq:P2}) on the left-hand side of
Eq.(\ref{eq:P1}), the result of substitution is expanded in the powers of
$\varepsilon$. Subsequently, the coefficients at the same power of
$\varepsilon$ are equated, leading to the following system of equations:
\begin{equation}
\dot{C}_{0}^{(0)}(\bm{\sigma},\bm{q})+i\tilde{W}_{0}(\bm{q}%
)C_{0}^{(0)}(\bm{\sigma},\bm{q})=0 \label{eq:P3}%
\end{equation}%
\begin{equation}
\dot{C}_{0}^{(1)}(\bm{\sigma},\bm{q})+i\tilde{W}_{0}(\bm{q}%
)C_{0}^{(1)}(\bm{\sigma},\bm{q})=f\left(  C^{(0)}\right)
\label{eq:P4}%
\end{equation}
\begin{equation}
\dot{C}_{0}^{(2)}(\bm{\sigma},\bm{q})+i\tilde{W}_{0}(\bm{q}%
)C_{0}^{(2)}(\bm{\sigma},\bm{q})=f_{C}'\left(
C^{(0)}\right)  C^{(1)} \label{eq:P5}%
\end{equation}
where $f_{C}'$ denotes the derivative with respect to $C$.

It can be easily confirmed that Eqs.(\ref{eq:P3}) and (\ref{eq:P4}) coincide
with the corresponding equations in Section \ref{sec:FrohlichPT}. Because
$C_{0}^{(1)}(\bm{\sigma},\bm{q})=0$, the right-hand side of
Eq.(\ref{eq:P5}) is also zero, resulting in $C_{0}^{(2)}(\bm{\sigma
},\bm{q})=0$. Consequently, higher-order coefficients are zero.


\end{document}